\documentclass[preprintnumbers, floatfix, letterpaper, onecolumn,aps,prd,epsfig,nofootinbib,natbib,longbibliography]{revtex4-2}
\usepackage{graphicx}
\usepackage{epstopdf}
\usepackage{latexsym}
\usepackage{amssymb}
\usepackage{amsmath}
\usepackage{color}
\usepackage{mathrsfs}
\usepackage{xparse}
\usepackage{float}
\usepackage{mathtools}
\usepackage[center]{subfigure}
\usepackage{placeins}
\usepackage{multirow}
\usepackage{tabularx}
\usepackage{float}
\usepackage{booktabs}
\usepackage{bm,graphicx,dcolumn,epstopdf,epsf, latexsym,mathbbol, amssymb,amsmath,color,slashed, mathrsfs,mathcomp,simplewick}
\usepackage[center]{subfigure}
\usepackage{multirow}
\usepackage{makecell}
\usepackage{epstopdf}
\usepackage[colorlinks,linkcolor=blue,citecolor=blue,urlcolor=blue]{hyperref}
\usepackage{hyperref}
\usepackage{amsmath}
\begin{document}
\renewcommand\arraystretch{2}
\newcommand{\bq}{\begin{equation}}
\newcommand{\eq}{\end{equation}}
\newcommand{\bqn}{\begin{eqnarray}}
\newcommand{\eqn}{\end{eqnarray}}
\newcommand{\nb}{\nonumber}
\newcommand{\lb}{\label}
\newcommand{\cb}{\color{blue}}
\newcommand{\cc}{\color{cyan}}
\newcommand{\cm}{\color{magenta}}
\newcommand{\rc}{\rho^{\scriptscriptstyle{\mathrm{I}}}_c}
\newcommand{\rd}{\rho^{\scriptscriptstyle{\mathrm{II}}}_c} 
\NewDocumentCommand{\evalat}{sO{\big}mm}{%
  \IfBooleanTF{#1}
   {\mleft. #3 \mright|_{#4}}
   {#3#2|_{#4}}%
}

\newcommand{\PRL}{Phys. Rev. Lett.}
\newcommand{\PL}{Phys. Lett.}
\newcommand{\PR}{Phys. Rev.}
\newcommand{\CQG}{Class. Quantum Grav.}
\newcommand{\parallelsum}{\mathbin{\!/\mkern-5mu/\!}}
\renewcommand\arraystretch{2}
 \newcommand{\subbq}{\begin{subequations}}
 \newcommand{\subeq}{\end{subequations}}
\newcommand{\rcone}{\rho^{\scriptscriptstyle{\mathrm{I}}}_c}
\newcommand{\La}{\Lambda}
\newcommand{\va}{\scriptscriptstyle}
\newcommand{\be}{\nopagebreak[3]\begin{equation}}
\newcommand{\ee}{\end{equation}}
\newcommand{\sign}{\text{sign}}

\newcommand{\ba}{\nopagebreak[3]\begin{eqnarray}}
\newcommand{\ea}{\end{eqnarray}}

\newcommand{\la}{\label}
\newcommand{\n}{\nonumber}
\newcommand{\su}{\mathfrak{su}}
\newcommand{\SU}{\mathrm{SU}}
\newcommand{\U}{\mathrm{U}}

\def\be{\nopagebreak[3]\begin{equation}}
\def\ee{\end{equation}}
\def\ba{\nopagebreak[3]\begin{eqnarray}}
\def\ea{\end{eqnarray}}
\newcommand{\f}{\frac}
\def\rmd{\rm d}
\def\lp{\ell_{\rm Pl}}
\def\d{{\rm d}}
\def\fe{\mathring{e}^{\,i}_a}
\def\fw{\mathring{\omega}^{\,a}_i}
\def\fq{\mathring{q}_{ab}}
\def\t{\tilde}

\def\db{\delta_b}
\def\dc{\delta_c}
\def\T{\mathcal{T}}
%%THESE ARE NEW MACROS USED IN APPENDIX A
\def\GammaE{\Gamma_{\rm ext}}
\def\GammaEb{\bar\Gamma_{\rm ext}}
\def\GammaEh{\hat\Gamma_{\rm ext}}
\def\Hee{H_{\rm eff}^{\rm ext}}
\def\H{\mathcal{H}}

\newcommand{\R}{\mathbb{R}}
\providecommand{\apjs}{ApJS}

%%%%%%%%%%%%%%%%%%%%%%%%%%%%%%%%%%%%%%%%%%%%%%%%%%%%%%%%%%%%%%%%%%%%%%%%%%%%%%%%%%%%%%%%%%%%%%%%%%%

\title{Effects of the ekpyrotic mechanism on  inflationary phase  in mLQC-II} 

\author{Jared Fier$^{a}$}
\email{Jared$\_$Fier@baylor.edu}

\author{Aqsa Fareed$^{a}$}
\email{Aqsa$\_$Fareed1@baylor.edu}

\author{Reena Danika Bendukuri$^{b}$}
\email{reenadanikabendukuri@gmail.com}

\author{Anzhong Wang$^{a}$ \footnote{The corresponding author}
% \orcid{0000-0002-8852-9966}
}
\email{anzhong$\_$wang@baylor.edu}

\affiliation{$^{a}$ GCAP-CASPER, Department of Physics and Astronomy, Baylor University, Waco, TX 76798-7316, USA\\
$^{b}$ Adlai E. Stevenson High School, 1 Stevenson Drive, Lincolnshire, IL 60069, USA}

\date{\today}
\begin{abstract}
Bouncing cosmological models replace the big bang singularity with a regular bounce, but face a well-known obstacle: the anisotropic shear grows as $a^{-6}$ during contraction, where $a$ is the average expansion factor, threatening to dominate before a homogeneous, isotropic universe can emerge from the bounce. A standard remedy is a scalar field with an ekpyrotic-like potential that turns negative near the bounce, pushing its effective equation of state above unity so it outgrows the shear. Here we examine how this mechanism affects the subsequent inflationary phase within mLQC-II, a modified loop quantum cosmology (LQC) model in which inflation is otherwise generic. Taking a potential combining ekpyrotic and inflationary components, we numerically evolve the dynamical equations across the bounce for various parameter choices. We find that while some choices of the free parameters of the theory allow the ekpyrotic potential to dominate near the bounce and resolve the shear problem, it substantially reshapes the post-bounce evolution. In particular, parameter choices that yield sufficient inflation without the ekpyrotic mechanism now often fail to do so. Sufficiently long inflation remains possible but appears to require fine-tuning. As our results are numerical, a more systematic analysis will be needed to establish their generality.

\end{abstract}

\maketitle

\section{
Introduction
}
\renewcommand{\theequation}{1.\arabic{equation}}
\setcounter{equation}{0}

Ever since it was introduced in 1980 \cite{1981PhRvD..23..347G}, cosmic inflation has proven remarkably successful: it accounts for several persistent puzzles of the standard hot big bang model and remains in agreement with every cosmological and astrophysical dataset gathered to date \cite{Planck:2018jri,AtacamaCosmologyTelescope:2025blo}. Even so, the paradigm is not without difficulties. A prominent one is its acute sensitivity to ultraviolet (UV) physics, so that whether inflation truly succeeds depends on how that short-distance regime is understood \cite{Brandenberger:2012aj,Silverstein:2016ggb,Baumann:2014nda}. Concretely, whenever the inflationary era extends much beyond the minimal number of e-folds needed to address the classic problems, the comoving scales observed today can be traced back to modes whose physical wavelengths were sub-Planckian during inflation. In that situation, quantizing matter fields on a fixed classical geometry is no longer justified, since quantum-geometric effects should be substantial and a smooth classical spacetime description breaks down. This is the well-known trans-Planckian problem of cosmological perturbations \cite{Brandenberger:2012aj}. 

A related difficulty stems from the big bang singularity itself \cite{Borde:1993xh,Borde:2001nh}, where the very prescription of initial data becomes ambiguous. 
The usual workaround is to disregard the pre-inflationary epoch altogether and to fix initial conditions early enough that every observable mode lies well within the Hubble radius. Because slow-roll inflation drives the geometry toward a near–de Sitter state, the Bunch–Davies (BD) vacuum then presents itself as the canonical choice \cite{Bunch:1978yq}. Yet how such a state could arise dynamically within quantum cosmology (QC) remains unsettled, given that a pre-inflationary stage necessarily bridges the Planck and inflationary regimes—energy densities separated by roughly twelve orders of magnitude—during which particle production cannot be avoided. 
All of these open issues ultimately point back to QC, an area investigated intensively over recent decades through a variety of proposals, among them constructions rooted in string/M-theory \cite{Green_Schwarz_Witten_2012,Becker:2006dvp} and in loop quantum gravity (LQG) \cite{Ashtekar:2004eh,Thiemann_2007,Bojowald_2010,Gambini:2011zz,Rovelli:2014ssa}. 

Over the past twenty five years in particular, LQG techniques have been applied systematically to singularity resolution across a range of cosmological settings (for recent surveys, see Refs. \cite{Ashtekar:2011ni,ElizagaNavascues:2020uyf,Li:2023dwy,Agullo:2023rqq}), yielding a consistent portrait of Planck-scale dynamics: the big bang singularity is superseded by a quantum bounce driven entirely by quantum-geometric effects. The resulting framework is commonly known as loop quantum cosmology (LQC). Because the Hamiltonian constraint can be regularized in more than one way, several modified loop quantum cosmological (mLQC) models have been put forward to probe these ambiguities \cite{Li:2021mop}. Two have been studied most thoroughly, mLQC-I and mLQC-II \cite{Yang:2009fp}, distinguished by how the Lorentzian piece of the gravitational Hamiltonian constraint is treated relative to the Euclidean piece \cite{Li:2018opr,Li:2018fco,Li:2019ipm}. The present work centers on mLQC-II, which follows from the alternative quantization first put forward by Yang, Ding, and Ma \cite{Yang:2009fp}. In contrast to mLQC-I \cite{Yang:2009fp,Li:2021mop}, whose contracting branch settles into a quasi–de Sitter phase and which additionally admits a top-down derivation directly from LQG via Thiemann's regularization \cite{Assanioussi:2018hee,Assanioussi:2019iye,Dapor:2017rwv,Dapor:2017gdk,Han:2021cwb}, mLQC-II recovers classical general relativity on both sides of the bounce, rendering its background evolution qualitatively much closer to that of standard LQC. It is precisely this similarity with LQC that motivates a dedicated analysis of mLQC-II here. 

 In every bouncing scenario, be it classical \cite{Lehners:2008vx,Battefeld:2014uga,Brandenberger:2016vhg} or quantum \cite{Ashtekar:2011ni,Li:2023dwy,Agullo:2023rqq}, a persistent obstacle is the shear problem. During contraction the shear scales as $a^{-6}$, where $a$ represents the average scale factor, outpacing every matter component apart from a stiff fluid (equivalently, a massless scalar), which redshifts at the same rate. Even then it is far from obvious how the stiff fluid can systematically overtake the shear so as to leave behind a homogeneous, isotropic universe once the bounce is crossed. Shear dynamics in homogeneous anisotropic Bianchi models has been examined extensively within LQC \cite{Ashtekar:2011ni,Li:2023dwy,Agullo:2023rqq}, producing a number of notable results; for the Bianchi I case, in particular, the shear was shown to be asymptotically conserved \cite{Chiou:2007sp,Ashtekar:2009vc}. A standard remedy is therefore to invoke the ekpyrotic mechanism (see, e.g., \cite{McNamara:2022dmf,Motaharfar:2023hil} and references therein), originally devised in the context of colliding branes \cite{Khoury:2001wf,Lehners:2008vx} and later carried over to other bouncing frameworks, including matter and ekpyrotic/cyclic bounces \cite{Cai:2012va,Brandenberger:2016vhg,Ijjas:2019pyf,Ijjas:2020dws,Ijjas:2021zyf,Tukhashvili:2023itb,Ijjas:2024oqn,Itzhaki:2025gdv}. The idea is to add a scalar field carrying an ekpyrotic-like potential that turns negative around the bounce, pushing the field's effective equation of state (EoS) above unity so that its energy density climbs as $\rho_{\phi}\propto a^{-3(1+w)}$, where $w>1$, and thereby overwhelms the shear near the bounce. A homogeneous and isotropic universe can then emerge in the expanding phase.

 In the present paper we investigate how the ekpyrotic mechanism impacts the inflationary epoch specifically within mLQC-II. The analogous question was addressed for LQC and mLQC-I in  \cite{Brown:2025hcb}. Since mLQC-II shares many background features with LQC while differing in its Planck-scale dynamics, it is natural to ask whether the conclusions reached there carry over, or whether the closer resemblance to LQC leads to quantitatively distinct behavior. This is especially pertinent because inflation is known to be generic in mLQC-II in the absence of any ekpyrotic potential \cite{Li:2019ipm}.  The central question is thus whether that generality survives once the ekpyrotic mechanism is switched on. To address it, we take a single scalar field governed by the total potential,
 \begin{equation}
     \lb{eq1.1}
     V(\phi)=V_{\text{inf}}(\phi)+V_{\text{ek}}(\phi),
 \end{equation}
 in which $V_{\text{ek}}(\phi)$ is of ekpyrotic potential and $V_{\text{inf}}(\phi)$ is the inflationary potential. For the mechanism to operate, $V_{\text{ek}}(\phi)$ must control the contracting evolution in the vicinity of the bounce, whereas after the bounce $V_{\text{inf}}(\phi)$ should progressively grow and eventually take over, ushering in an inflationary stage. The problem then becomes one of demonstrating that this hand-off actually takes place for a suitable set of initial data, and that the ensuing inflation lasts long enough to resolve the big bang puzzles that motivated inflation to begin with \cite{1981PhRvD..23..347G}. We stress that in matter and ekpyrotic bounces \cite{Lehners:2008vx,Battefeld:2014uga,Brandenberger:2016vhg,Cai:2012va,Ijjas:2019pyf,Ijjas:2020dws,Ijjas:2021zyf,Tukhashvili:2023itb,Ijjas:2024oqn,Itzhaki:2025gdv} an inflationary phase is not assumed. Indeed a chief aim of those constructions is to replace inflation entirely by letting a regular bounce take care of the singularity and trans-Planckian problems, while a matter-dominated contraction additionally generates a scale-invariant spectrum \cite{Wands:1998yp}. Along these lines, the inflation-free primordial spectrum has also been explored in LQC and mLQC-I \cite{Bojowald:2004kt,Wilson-Ewing:2012lmx,Wilson-Ewing:2013bla,Wilson-Ewing:2015sfx,Li:2021fmu}, though it was recently shown to be at odds with current data \cite{Li:2020pww}. For this reason we concentrate on mLQC-II, a model that retains an inflationary phase which, as noted above, is generic there once the ekpyrotic mechanism is absent \cite{Li:2019ipm}. Scanning a broad range of cases numerically, we find that, with an appropriate choice of the free parameters, the ekpyrotic-like potential does dominate through the bounce region, where the scalar-field EoS exceeds unity and the shear problem is accordingly resolved. Sufficiently after the bounce the inflationary potential takes command, and an inflationary era develops that can be prolonged enough to cure the standard-cosmology problems. 
 %However, to realize the above claims, it seems that fine-tuning is required for the choices of the initial conditions. 
 
 The remainder of the paper is organized as follows. Section II provides a concise account of mLQC-II together with the relevant Hamiltonian equations of motion. In Sec. III we integrate these equations numerically for the total potential of Eq.(\ref{eq1.1}) across various parameter choices within mLQC-II, while considering the effects of inflation with and without an ekpyrotic mechanism. Although whether a sufficiently long inflationary phase arises depends sensitively on the parameters, we do identify regions of parameter space of nonzero measure that yield the desired inflation, albeit with indications that some fine-tuning may be needed. Section IV summarizes our findings and offers concluding remarks.
%%%%%%%%%%%%%%%%%%%%%%%%%%%%%%%%%%%%%%%%%%%%%%%%%%%%%%%%%%

\section{Effective Dynamical Equations in mLQC-II}
\renewcommand{\theequation}{2.\arabic{equation}}
\setcounter{equation}{0}
\lb{SecII}
%%%%%%%%%%%%%%%%%%%%%%%%%%%%%%%
%%%%%%%%%%%%%%%%%%%%%%%%%%%%%%

In this section, we provide a summary of the modified Friedmann dynamics in the framework of mLQC-II \cite{Li:2018fco}. Its dynamics can be obtained from the effective Hamiltonian given by,
\begin{equation}
\lb{eq2.1}
\mathcal{H} = - \frac{3v}{2\pi G \lambda^2 \gamma^2 }\sin^2 \left(\frac{\lambda b}{2}\right)\left[1+\gamma^2\sin^2{\left(\frac{\lambda b}{2}\right)} \right] +\mathcal{H_M},
\end{equation}
where  $G$ is the Newtonian constant, $v \equiv a^3$,  
and $a$ is the expansion factor of the Universe,
\begin{equation}
\lb{eq2.2}
ds^2 = - dt^2 + a^2(t)\left(dx^2 + dy^2 +dz^2\right).
\end{equation}
The variable $b$ denotes the momentum conjugate of $v$ and obey the Poisson bracket
\begin{equation}
\lb{eq2.3}
\{b,v\} = 4 \pi G \gamma,
\end{equation}
where  $\gamma$ is known as the Barbero-Immirzi parameter whose value is set to $\gamma \approx 0.2375$ using black hole thermodynamics in LQG \cite{Meissner:2004ju}. The parameter $\lambda$ is defined as $\lambda^2 \equiv \Delta = 4 \sqrt{3}\pi\gamma\ell^2_\text{pl}$, where $\Delta$ denotes the minimal non-zero area gap of the area operator in LQG \cite{Ashtekar:2004eh,Thiemann_2007,Bojowald_2010,Gambini:2011zz,Rovelli_Vidotto_2014,Ashtekar:2017yom}. The matter 
Hamiltonian $\mathcal{H_M}$ is given by
\begin{equation}
\lb{eq2.4}
\mathcal{H_M} = v \rho,  
\end{equation}
where $\rho$ denotes the energy density of the matter field. Then, the Hamiltonian equation for a given physical quantity $A$ of the system
\begin{equation}
\lb{eq2.5}
\dot{A} = \left\{A, \mathcal{H}\right\},   
\end{equation}
yields
\bqn
\lb{eq2.6}
\dot{b} &=& \left\{b, \mathcal{H}\right\} =  4 \pi G \gamma \frac{\partial \mathcal{H}}{\partial v}   
= -\frac{6\sin^2{\left(\frac{\lambda b}{2}\right)}}{\gamma \lambda^2}\left[1+\gamma^2\sin^2{\left(\frac{\lambda b}{2}\right)}\right]-4\pi G\gamma P,\\
\lb{eq2.7}
\dot{v} &=& \left\{v, \mathcal{H}\right\} = - 4 \pi G \gamma \frac{\partial \mathcal{H}}{\partial b}   
= \frac{3v\sin{(\lambda b)}}{\gamma \lambda}\left[1+\gamma^2 -\gamma^2\cos{(\lambda b)} \right],
\eqn 
where
 \begin{equation}
\lb{eq2.11}
 P \equiv -  \frac{\partial \mathcal{H_M}}{\partial v}. 
 \end{equation}

When we consider a scalar field $\phi$ with a potential $V(\phi)$, we have 
\begin{equation}
\lb{eq2.12}
\mathcal{H_{\phi}} = \frac{p_{\phi}^2}{2v} + v V(\phi),  
\end{equation}
where $p_{\phi}$ is the momentum conjugate of $\phi$ and satisfies the canonical relation 
\begin{equation}
\lb{eq2.13}
\{\phi,p_{\phi}\} = 1. 
\end{equation}
Then, the Hamiltonian equation (\ref{eq2.5}) yields
\bqn
\lb{eq2.14}
\dot \phi &=& \{\phi,\mathcal{H}\} = \frac{\partial\mathcal{H_{\phi}}}{\partial p_{\phi}} = \frac{p_{\phi}}{v}, \\
\lb{eq2.15}
\dot{p}_{\phi} &=& \{p_{\phi},\mathcal{H}\} = -\frac{\partial\mathcal{H_{\phi}}}{\partial \phi} = - v V_{,\phi},
\eqn
where $V_{,\phi} \equiv dV(\phi)/d\phi$. From the above equations, we are able to produce the EoM for the scalar field,
\bqn
\lb{eq2.16}
\ddot\phi + 3H \dot\phi + V_{,\phi}(\phi) = 0,  
\eqn
which reproduces the Klein-Gordon equation.

By applying the Hamiltonian constraint $\mathcal{H} \simeq 0$ we find that 
 \begin{equation}
    \lb{eq2.8}
 \rho   =  \frac{3}{2\pi G\lambda^2 \gamma^2} \sin^2{\left(\frac{\lambda b}{2}\right)}\left[1+\gamma^2\sin^2{\left(\frac{\lambda b}{2}\right)} \right]. 
 \end{equation}

One can invert the density function to obtain,
\begin{equation}
     \lb{eq2.8.3}
    \sin^2(\lambda b_{\pm}/2) = \frac{-1\pm \sqrt{1+\gamma^2 \rho/\rho_c^{\text{II}}}}{2\gamma^2},
\end{equation}
 where  the critical density  $\rho_c^{\text{II}}$ in mLQC-II is defined as
  \begin{equation}
      \lb{eq2.8.2}
      \rho_c^{\text{II}} = \frac{3}{8\pi G \lambda^2 \gamma^2}.
  \end{equation}
It can be shown that $b_+$ represents the only physical root since $b$ must be real \cite{Li:2018opr,Li:2018fco,Li:2019ipm}.  Substituting the above expression into Eqs.(\ref{eq2.6}) and (\ref{eq2.7}) we find the equations
 \begin{align}
 \lb{eq2.9}
 \dot{b} &=  -  4 \pi G \gamma \left(\rho + P\right),  \\
 \lb{eq2.10}
 H^2
&=\frac{16\pi G \rho}{3}\left(1-\frac{\rho}{\rho^{\scriptscriptstyle{\mathrm{II}}}_c}\right)\left(\frac{1+4\gamma^2(\gamma^2+1)\rho/\rho^{\scriptscriptstyle{\mathrm{II}}}_c}{1+2\gamma^2\rho/\rho^{\scriptscriptstyle{\mathrm{II}}}_c+\sqrt{1+4\gamma^2(1+\gamma^2)\rho/\rho^{\scriptscriptstyle{\mathrm{II}}}_c}}\right).   
 \end{align}

Reading off the energy density and pressure from  Eqs.(\ref{eq2.4}), (\ref{eq2.11}) and (\ref{eq2.14}) we find that
\bqn
\lb{eq2.17}
\rho_{\phi} =  \frac{p_{\phi}^2}{2v^2} +  V(\phi) = \frac{1}{2}\dot{\phi}^2 + V(\phi),\quad
P_{\phi} =  \frac{p_{\phi}^2}{2v^2} -  V(\phi) = \frac{1}{2}\dot{\phi}^2 - V(\phi).
\eqn
The equation of state (EoS) for the scalar field is then given by
\bqn
\lb{eq2.18}
w_{\phi} \equiv \frac{P_{\phi}}{\rho_{\phi}} =   \frac{\frac{1}{2}\dot{\phi}^2 - V(\phi)}{\frac{1}{2}\dot{\phi}^2 + V(\phi)} =
\begin{cases}
    \geq 1, & V(\phi) \leq 0, \cr
     \leq 1, & V(\phi) \geq 0, \cr
\end{cases}, ~~~
\eqn
provided that $\rho_{\phi} > 0$.  We can see that the sign of the potential will then determine the nature of the field as a fluid which will become a key point in how the ekpyrotic mechanism will suppress the shear generated at the bounce.

Eqs.(\ref{eq2.6}), (\ref{eq2.7}),  (\ref{eq2.14}) and (\ref{eq2.15}) are the first-order ordinary differential equations for the four canonical variables ($v, b; \phi, p_{\phi}$). Once the initial conditions are specified at a given moment,  they uniquely determine the trajectory of the evolution of the Universe. Such initial conditions are often imposed at the quantum bounce \cite{Ashtekar:2011ni,Li:2021mop}, at which the expansion factor reaches its minimal value and the energy density reaches its maximum.

In addition, the advantage of imposing the initial conditions at the bounce is that the time derivative of the scalar field at the bounce $\dot\phi_B$ is determined uniquely up to a sign for any given initial scalar field value at the bounce $\phi_B$ via the relation $\rho(t_B) = \rho_c^{\text{II}}$, where $t_B$ denotes the time of the bounce,  
which yields
\begin{equation}
\lb{eq2.19}
\dot\phi_B = \pm\sqrt{2(\rho_c^{\text{II}}-V(\phi_B))}.
\end{equation}  
Furthermore, these equations are invariant under the rescaling $a \rightarrow a/L_o$, which allows us to fix $a_B = 1$ at the time of the bounce with no loss of generality.  Then, the initial conditions are reduced to the choice of
\begin{equation}
\lb{eq2.20}
\left(\phi_B, \text{sgn}\left(\dot\phi_B\right)\right).
\end{equation}
Moreover, using the translation invariance $t \rightarrow t + t_0$, in the rest of this paper, we shall set $t_B = 0$.

%%%%%%%%%%%%%%%%%%%%%%%%%%%%%%%%%%%%%%%%%%%%%%%%%%%%%%%%%%

\section{Effects of Ekpyrotic Mechanism on Inflation in mLQC-II}
\renewcommand{\theequation}{3.\arabic{equation}}
\setcounter{equation}{0}
\lb{SecIII}

It is well-known that shear behave like a stiff fluid  \cite{Ryan:1975jw}
\bq
\lb{eq3.1}
\sigma^2 \equiv \sigma_{\mu\nu}\sigma^{\mu\nu} = \frac{\Sigma^2}{a^6},
\eq
where $a(t)$ is the average expansion factor of the 3-volume of a universe, $\Sigma^2$ is a constant, and $\sigma_{\mu\nu}$ denotes the anisotropic shear tensor, defined via the relation
\bqn
\lb{eq3.1aa}
\nabla_{\nu}v_{\mu} = \frac{1}{3}\left(g_{\mu\nu} + v_{\mu}v_{\nu}\right)\theta + \omega_{\mu\nu} + \sigma_{\mu\nu}.
\eqn
Here $v^{\mu}$ denotes the unit tangential vector of the time-like geodesics, $\theta$ and $\omega_{\mu\nu}$  denote respectively the expansion scalar and vorticity  tensor of the time-like geodesics. In the homogeneous universe, we have  $\omega_{\mu\nu} = 0$ and $\theta = 3H$.   

 For the kinetic energy dominated initial conditions, the massless scalar field also behaves like a stiff fluid, so we have 
\bq
\lb{eq3.2}
\rho_{\phi} \simeq P_{\phi} = \frac{\rho^{(0)}_{\phi}}{a^6},\;\; (t \simeq t_B),
\eq
where   $\rho^{(0)}_{\phi}$ is a constant. 
Therefore, it is not always clear which one shall dominate the evolution of the universe near the bounce. If the shear dominates, the universe will become highly anisotropic after the bounce, whereby a homogeneous and isotropic universe cannot be developed. Therefore, it is crucial for any bounce model, such as mLQC-II, to be considered as viable, one has  to make sure that the shear does not dominate in the contracting phase, especially near the bounce \cite{Lehners:2008vx,Battefeld:2014uga,Brandenberger:2016vhg}. 
One way   is to introduce the ekpyrotic potential \cite{Khoury:2001wf,Cai:2012va,Motaharfar:2023hil}
 \begin{equation}
 \lb{eq3.3}
    V_{\text{ekp}}(\phi)=-\frac{2 U_0}{e^{-\sqrt{\frac{16 \pi}{p}}\phi}+e^{\beta\sqrt{\frac{16 \pi}{p}}\phi}},
\end{equation}
so that near the bounce we have $V(\phi_B) < 0$, where $U_0, \; p$ and $\beta$ are all positive and otherwise free parameters. 
Then, we have
$w_{\phi} > 1$, and
\begin{equation}
\lb{eq3.4}
\rho_{\phi}^{\text{ekp}}   \propto \frac{1}{a^{3(1+w_{\phi})}},\;\; (t \simeq t_B),
\end{equation}
so the scalar field will dominate the evolution of the universe and the effects of the shear will be suppressed. As a result, the contracting universe can smoothly evolve into an expanding  homogeneous and isotropic one. 

When far away from the bounce, we would expect to obtain an inflationary phase in the post-bounce region, $t \gg t_B$ \footnote{It should be noted that in most of the bouncing models,  the inflationary phase is not required, see, for example,   \cite{Lehners:2008vx,Battefeld:2014uga,Brandenberger:2016vhg}. This is fundamentally different from quantum bouncing models of LQG, in which it has been shown that inflation after the bounce is generic in LQC \cite{Ashtekar:2011rm} and mLQC-I and mLQC-II \cite{Li:2019ipm}.}. This is possible if the total potential $V(\phi)$ consists of two parts
\begin{equation}
 \label{eq3.5}
    V(\phi)=V_\text{ekp}(\phi)+V_{\text{inf}}(\phi),
\end{equation}
where   $V_{\text{inf}}(\phi)$ denotes an inflationary potential and will dominate the evolution of the universe when $t \gg t_B$, while for $t \simeq t_B$ the ekpyrotic potential $V_{\text{ekp}}(\phi)$  dominates. 

Following Planck 2018 data \cite{Planck:2018jri}, inflation with various known potentials have been ruled out, including   potentials with the form $V(\phi) \propto \phi^n$.  However, models with polynomial chaotic potentials can fit the observations well \cite{Destri:2007pv,Nakayama:2013jka,Kallosh:2014xwa}. A typical example is \cite{Kallosh:2025ijd}
\begin{equation}
\label{eq3.11}
V_{\text{inf}}(\phi) = \frac{1}{2}m^2\phi^2\left(1 - \alpha_1\phi + \alpha_2\phi^2\right)^2,  
\end{equation}
where $\alpha_{1, 2}$ are two coupling constants. By properly choosing these constants, it can be shown that the models
fit the observational data very well. In particular, the choice of  $\alpha_1 = 0.14$ and $\alpha_2 = 6.644\times 10^{-3}$
allows the model to fit very well to the current Atacama Cosmology Telescope (ACT) observations \cite{ACT:2025fju}.

In this paper, we shall consider the polynomial chaotic potentials given above as a representative case, and the generalization of our analysis to other viable potentials are straightforwards.

Then, a natural question is whether or not a mechanism mentioned above exists. Our following  analysis shows that this can indeed be the case by properly choosing the parameters involved in the models, despite the fact that the effects of the ekpyrotic-like potential are dramatic. 

\begin{figure}[h!]
\includegraphics[width=0.45\linewidth]{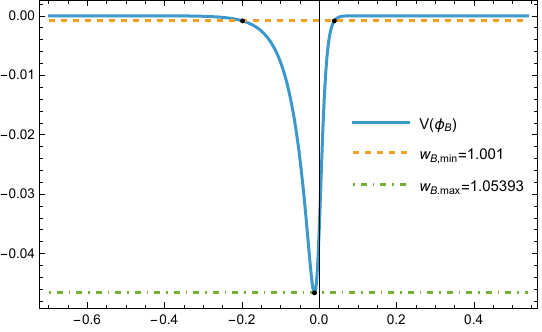}
\caption{The plot of the total potential $V(\phi)$ defined by Eq.(\ref{eq3.5}) with $U_0=0.0366$, $p=0.1$, $\beta=5$ and the chaotic inflationary potential given by Eq.(\ref{eq3.11}) with $\alpha_1 = \alpha_2 = 0,\; m = 1.23 \times 10^{-6}\; m_{P}$. The corresponding  minimal and maximal values of $w_B$ are also given. Here $m_p \equiv 1/(8\pi G)^{1/2}$.}
\label{fig1}
\end{figure}

\begin{figure}[h!]
\includegraphics[width=0.45\linewidth]{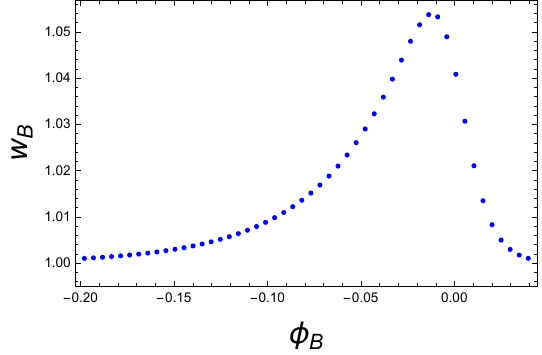}
\caption{The plot of the $w_B$ for different initial $\phi_B$ values with $U_0=0.0366$, $p=0.1$, $\beta=5$ and the chaotic inflationary potential given by Eq.(\ref{eq3.11}) with $\alpha_1 = \alpha_2 = 0,\; m = 1.23 \times 10^{-6}\; m_{P}$.}
\label{fig2}
\end{figure}

For our above claim, let us first show how to choose the initial conditions at the bounce $t = t_B$ with a total potential given by Eq.(\ref{eq3.5}). First, from Eqs. (\ref{eq2.18}) and (\ref{eq2.19}) we find  
\begin{equation}
    \label{eq3.6}
    V(\phi_B)= - \frac{w_B - 1}{2}\rho_c^{\text{II}},
\end{equation}
where $w_B \equiv w_{\phi}(\phi_B)$. For any given potential $V(\phi)$ and a fixed equation of state $w_B > 1$, we can solve the above equation for $\phi_B$.  In particular, starting with a minimal value of $w_B$, say, $w_{\text{Bmin}}=1.001$,   we can solve Eq.(\ref{eq3.6}) numerically to obtain the corresponding range of $\phi_B$.   As shown in Fig.\ \ref{fig1}, the maximal value of $w_{\text{Bmax}}$ is obtained when the potential is at its minimum  $V_{\text{min}}(\phi_B)$ with
\begin{equation}
    \label{eq3.6a}
    V_{\text{min}}(\phi_{B}) = - \frac{w_{\text{Bmax}} - 1}{2}\rho_B,
\end{equation} 
denoted by the crossing point of the horizontal straight line $- (w_{\text{Bmax}} - 1)\rho_B/2$ and $V(\phi_B)$.  The full range of $w_B$ values within the allowed $\phi_B$ values is show in Fig.{\ref{fig2}}.

 %%%%%%%%%%%%%%%%%%%%%%%%%%%%%%%%%%%%%%%%%%%%%%%%%%%%%%%%%%%%%%%%%%%%%%%%%%%%%%%%%%%%%%%%%%%%%%%%%%%% Figure 1
%%%%%%%%%%%%%%%%%%%%%%%%%%%%%%%%%%%%%%%%%%%%%%%%%%%%%%%%%%%%%%%%%%%%%%%%%%%%%%%%%%%%%%%%%%%%%%%%%%%%%%%%%%%%%%%%
%%%%%%%%%%%%%%%%%%%%%%%%%%%%%%%%%%%%%%%%%%%%%%%%%%%%%%%%%%%%%%%%%%%%%%%%%%%% 

%%%%%%%%%%%%%%%%%%%%%%%%%%%%%%%%%%%%%%%%%%%%%%%%%%%%%%%%%%%%%%%%%%%%%%%%%%%%%%%%%%%%%%%%%%%%%%%%%%%% Figure 2 
%%%%%%%%%%%%%%%%%%%%%%%%%%%%%%%%%%%%%%%%%%%%%%%%%%%%%%%%%%%%%%%%%%%%%%%%%%%%%%%%%%%%%%%%%%%%%%%%%%%%%%%%%%%%%%%%
%Figure2%%%%%%%%%%%%%%%%%%%%%%%%%%%%%%%%%%%%%%%%%%%%%%%%%%%%%%%%%%%%%%%%%%%%%%%%%%% 

With the above chosen initial conditions for $\phi_B$, we can study the evolution of the universe for any given   inflationary potential. In particular, let us first introduce  the quantities
 \cite{Baumann:2009ds}
\bqn
\lb{eq3.11a}
\epsilon_H = -\frac{\dot H}{H^2}, \quad \eta_H =  \frac{\ddot H}{2H \dot H}, \quad
%\lb{eq3.11b}
\epsilon_V = \frac{1}{16\pi G}\left( \frac{V_{,\phi}}{V}\right)^2,  \quad \eta_V = \frac{V_{,\phi\phi}}{8\pi G V}.
\eqn 
Then, we find that  
\begin{equation}\label{a and H relation}
\frac{\ddot a}{a} = H^2(1-\epsilon_H).
\end{equation}
The Universe experiences an accelerated expansion whenever $\epsilon_H < 1$, whereas slow-roll inflation occurs only when \cite{Baumann:2009ds}
\begin{equation}
\lb{eq3.14a}
\epsilon_H(t), \;\; \left|\eta_H(t)\right| \ll 1. 
\end{equation}
For the sake of concreteness, we define the onset of inflation as the time $t_i$ when $\epsilon_H(t_i) = 1$ for the first time in the transition phase, where $\epsilon_H < 1$ for $t > t_i$. The end of the inflationary phase is defined at the time $t_{\text{end}}$ when $\epsilon_H(t_{\text{end}}) = 1$ again for the first time after $t_i$. Therefore, for $t \in (t_i, t_{\text{end}})$ we have $\epsilon_H < 1$ and $\ddot{a} > 0$, that is, the universe is in its inflationary phase.  

Certainly, for the inflationary phase to be slowly rolling, the conditions (\ref{eq3.14a}) need to be satisfied during the inflation. Once these conditions are satisfied, we have \cite{Yogesh:2024iip}
\begin{equation}
\lb{eq3.14b}
\epsilon_H \simeq \epsilon_V, \quad \eta_H \simeq \eta_V - \epsilon_V, \;\; (\epsilon_H, |\eta_H| \ll 1). 
\end{equation}

The e-fold $N_{\text{inf}}$ during the inflationary phase is  defined as
\begin{equation}
\label{efolds definition}
N_{\text{inf}}  = \ln\left({\frac{a(t_{\text{end}})}{a(t_i)}}\right).
\end{equation}
To have a successful inflation, the inflation potential has to be very flat, so that the Universe can expand large enough \cite{Baumann:2009ds}. All the cosmological problems can be resolved if the Universe expands about 60 e-folds during the inflationary phase, although its exact value depends on the inflationary models \cite{Planck:2018jri}. Therefore, in the following we shall require   $N_{\text{inf}} \gtrsim 60$, although our main conclusions do not depend on its precise value.  From the above definition it is clear that in general $N_{\text{inf}}$ depends on the specific value of $\phi_B$.

\subsection{Inflation Without the Ekpyrotic Mechanism}

To study the effects of the Ekpyrotic mechanism during inflation, we will review how to calculate the probability for inflation in mLQC-II without an ekpyrotic mechanism \cite{Li:2019ipm}. To begin with, let us consider the chaotic potential, which corresponds to Eq.(\ref{eq3.11}) with $\alpha_1=\alpha_2=0$, even though this has been ruled out by observations \cite{Planck:2018jri}.  From here, we can investigate the general behavior as these conclusions do not depend on a specific form of the potential, and then apply this to more favorable potentials like the polynomial chaotic potential where $\alpha_1=\alpha_2\neq0$.

For the inflationary sector described by Eq.(\ref{eq3.11}), the first task is to determine the mass parameter $m$ from observationally constrained quantities. Following  \cite{Yogesh:2024iip}, we use the measured values of the scalar amplitude $A_s(k_*)$ and scalar spectral index $n_s(k_*)$, evaluated at the horizon-crossing of the pivot mode. Here the symbol $*$ denotes quantities evaluated at the horizon-crossing, $k = aH$.  Under the slow-roll approximation (\ref{eq3.14a}), the background variables at horizon crossing satisfy
\begin{align}
\lb{eqA.38}
 & H_*^2 = \pi m_{\text{P}}^2 \frac{\epsilon_V^* A_s(k_*)}{{\cal{F}}(k_*)},\\
 \lb{eqA.39}
 &  \xi_{\text{nBD}}(k_*)  - 2\left(3\epsilon_V^* - \eta_V^*\right) = n_s(k_*) -1, \\
 \lb{eqA.40}
& H_*^2 \simeq   \frac{8\pi }{3 m_{\text{P}}^2 }V(V_0,\phi_*),
\end{align}
where we define the quantities $\mathcal{F}(k)$ and $\xi_{\text{nBD}}$ as,
%\begin{widetext}
\begin{equation}
\lb{eqA.43}
 {\cal{F}}(k) = 1 + \delta_{\text{PL}}(k), \quad   
 \xi_{\text{nBD}}(k_*)  \equiv  \left. \frac{d\ln{\cal{F}}(k)}{d\ln k}\right|_{k = k_*},
\end{equation}
where $\delta_{\text{PL}}(k)$ is purely due to quantum geometric effects \cite{Zhu:2017jew}. In classical theory we have  
$\delta_{\text{PL}}(k) = 0$. % \end{widetext}
The horizon-crossing conditions are supplemented by three additional relations \cite{Yogesh:2024iip},
\begin{align}
  \lb{eqA.41}
 & k_* = a_* H_*,\\
 \lb{eqA.42}
& \dot\phi_* \simeq - \frac{V_{,\phi}(V_0,\phi_*)}{3H_*},\\
 \lb{eqA.43}
 & a_B(a_*, \phi_*, \dot\phi_*) =  1. 
\end{align}
Together, Eqs.(\ref{eqA.38})-(\ref{eqA.40}) and Eqs.(\ref{eqA.41})-(\ref{eqA.43}) provide six independent constraints for the six quantities, 
$$
\left(a_*,H_*,\phi_*,\dot{\phi}_*,k_*,V_0\right).
$$
Consequently, once the observational values $(A_s,n_s)$ are specified, the horizon-crossing configuration is completely determined.  Moreover, knowledge of $(a_*,\phi_*,\dot{\phi}_*)$ fixes the subsequent and prior evolution through the modified Friedmann and Klein-Gordon equations, allowing the cosmological history to be reconstructed across both the contracting and expanding branches of the bounce \cite{Yogesh:2024iip}.

An important simplification arises because the pivot scale corresponds to an extremely large physical wavelength today. Since $k_*=a_0 e^{N_T}\simeq\mathcal{O}(m_P)$ with the total number of e-folds satisfying $(N_T \equiv ln(a_0/a_B)\gtrsim 141)$ \cite{Zhu:2017jew}, one finds that the quantum-geometry corrections entering the scalar spectrum are highly suppressed, $|\mathcal{F}(k_*)-1|$, $|\xi_{\text{nBD}}(k_*)|\ll 1$ \cite{Yogesh:2024iip} \footnote{It should be noted that in \cite{Yogesh:2024iip} only the LQC case was considered. However, from the results presented in \cite{Li:2021mop} it can be seen that this is also true in mLQC-II.}. These corrections can therefore be neglected, allowing us to take $\mathcal{F}(k)=1$, $\xi_{\text{nBD}}(k_*)=0$, and so simplifying Eqs.(\ref{eqA.38})-(\ref{eqA.40}) to \cite{Ashtekar:2011rm}
\begin{align}
\lb{eqA.38aa}
 & H_*^2 = \pi m_{\text{P}}^2 \epsilon_V^* A_s(k_*),\\
 \lb{eqA.39aa}
 &   2\left(3\epsilon_V^* - \eta_V^*\right) = 1 - n_s(k_*), \\
 \lb{eqA.40aa}
& H_*^2 \simeq   \frac{8\pi }{3 m_{\text{P}}^2 }V(V_0,\phi_*).
\end{align}

The parameter set $(A_s, n_s, \alpha_1, \alpha_2)$, together with Eqs.\eqref{eqA.39aa} and \eqref{eq3.11a}, determines the field value $\phi_*$. Substituting this result into Eq.~\eqref{eqA.38aa} yields the corresponding Hubble parameter at horizon crossing, $H_*$. The inflationary mass parameter $m$ is then obtained from Eqs.~\eqref{eq3.11a} and \eqref{eqA.40aa}. Once the triplet $(H_*, \phi_*, m)$ has been fixed, the remaining quantities $(a_*, \dot{\phi}_*, k_*)$ can be determined from Eqs.~\eqref{eqA.41}--\eqref{eqA.43} together with the dynamical equations.

For the probability analysis discussed below, the only quantity that must be inferred from observations is the mass parameter $m$. It is therefore sufficient to solve Eqs.~\eqref{eqA.38aa}--\eqref{eqA.40aa} for each choice of $(A_s, n_s, \alpha_1, \alpha_2)$. The observational inputs adopted in this work are taken from recent CMB measurements. For the pivot scale $k_*/a_0=0.05\,{\rm Mpc}^{-1}$, the Planck 2018 collaboration reported \cite{Planck:2018jri} 
\begin{equation} 
\label{Planck} 
\left(A_s,n_s\right)_{\rm Planck} = \left(2.0989\times10^{-9},\,0.9649\right), 
\end{equation} 
while the ACT analysis found \cite{AtacamaCosmologyTelescope:2025blo} 
\begin{equation}
\label{ACT} 
\left(A_s,n_s\right)_{\rm ACT} = \left(2.1179\times10^{-9},\,0.9666\right).
\end{equation} 
Combining the Planck 2018, ACT 2025, and DESI DR2 data sets yields \cite{AtacamaCosmologyTelescope:2025blo,DESI:2025zpo,DESI:2025zgx} 
\begin{equation} 
\label{Combined}
\left(A_s,n_s\right)_{\rm Combined} = \left(2.1370\times10^{-9},\,0.9752\right). 
\end{equation} 
This combined result is particularly noteworthy because it significantly constrains the viable inflationary parameter space, ruling out several inflationary scenarios that had previously been considered observationally consistent \cite{Kallosh:2025ijd}.

Solving Eqs.(\ref{eqA.38aa})-(\ref{eqA.40aa}) for each set of ($A_s, n_s$) given respectively by Eqs.(\ref{Planck}) (\ref{ACT}) and (\ref{Combined}), we find the solutions of $(H_*, \phi_*, m)$ for the chaotic potential $\alpha_1 = \alpha_2 = 0$, given in Table \ref{TableI}. It is interesting to note the significantly different values of $m$ with different observations. In particular, the combination of Planck and ACT leads to a value of $m$ that is significantly smaller than these obtained respectively by Planck 2018 and ACT 2025.

%%%%%%%%%Table I - Chaotic No Ekpyrotic
%%%%%%%%%%%%%%%%%%%%%%%%%%%%%%%%%%%%%%%%%%%%%%%%%%
%%%%%%%%%%%%%%%%%%%%%%%%%%%%%%%%%%%%%%%%%%%%%%%%%%

\begin{table*}[t]
%\label{TableI}
\centering
\begin{tabular}{lcccccccc}
         & $m$                  & $H_*$                & $\phi_*$ & $\phi_{\text{min}}$   & $\phi_{\text{max}}$  & $\phi_1$ & $\phi_2$ & $P$\\ \hline \hline
Planck   & $1.23\times 10^{-6}$ & $7.61\times 10^{-6}$ & $3.01$   & $-1.51\times 10^{-6}$ & $1.51\times 10^{-6}$ &          $-5.41$&          $0.72$&                  $2.59\times10^{-6}$  \\ \hline 
ACT      & $1.18\times10^{-6}$  & $7.45\times 10^{-6}$ & $3.09$   & $-1.58\times 10^{-6}$ & $1.58\times 10^{-6}$ &          $-5.41$&          $0.69$&                  $2.46\times10^{-6}$\\ \hline
Combined & $8.80\times 10^{-7}$ & $6.45\times10^{-6}$    & $3.58$   & $-2.11\times 10^{-6}$ & $2.11\times 10^{-6}$ &          $-5.48$&          $0.66$&                 
$1.85\times10^{-6}$ 
\end{tabular}
\caption{The physical quantities $(H_*,\phi_*,m)$ for a given set of $(A_s,n_s)$ from the observations, Planck, ACT or Planck + ACT + LB2 (Combined), and the corresponding $\phi_{\text{min}}$, $\phi_{\text{max}}$, $\phi_1$, $\phi_2$, and $P$ for $\dot{\phi}>0$ in a chaotic potential, $\alpha_1\alpha_2=0$ found in Eq.\eqref{eq3.11}.}
\label{TableI}
\end{table*}

%%%%%%%%%Table I  -  Chaotic no Ekpy
%%%%%%%%%%%%%%%%%%%%%%%%%%%%%%%%%%%%%%%%%%%%%%%%%%
%%%%%%%%%%%%%%%%%%%%%%%%%%%%%%%%%%%%%%%%%%%%%%%%%%

With the masses successfully solved for, our next task is to determine the range of possible $\phi$ values that are allowed.  Since $\rho\le\rho_c^{\text{II}}$, we find that $\phi$ must satisfy the condition $\frac{1}{2}\dot{\phi}^2+V(\phi)\le\rho_c^{\text{II}}$.  Then, the equation,
\begin{equation}
\lb{phiMN}
V_{\text{inf}}(\phi) = \rho_c^{\text{II}},
\end{equation}
determines the range of the scalar field $\phi$. For example, for the chaotic potential we find that 
\begin{equation}
\lb{eq3.16bb}
\phi \in \left(\phi_{\text{min}}, \phi_{\text{max}}\right), 
\end{equation}
where $\phi_{\text{max}}$ and $\phi_{\text{min}}$ are given in Table \ref{TableI}.  

To quantify how frequently a successful inflationary phase occurs, we employ the Liouville measure introduced in \cite{Ashtekar:2011rm}. For a given event $E$, the corresponding probability is defined by 
\begin{equation} 
\label{4.9} 
P(E)=\frac{1}{\mathcal D}\int_{\mathcal I(E)} d\omega , 
\end{equation} 
where the induced measure on the bounce surface is 
\begin{align} 
\label{4.9aa} 
d\omega &\equiv \left\{ -2\left[\hat{\mathcal H}^{\rm grav}(b_B)+V(\phi)\right] \right\}^{1/2} d\phi , \\ 
\hat{\mathcal H}^{\rm grav} &\equiv \frac{1}{v}\mathcal H^{\rm grav} = \frac{1}{v} \left( \mathcal H-\mathcal H_{\mathcal M} \right). 
\end{align} 
Here, $\mathcal I(E)$ denotes the subset of the bounce initial-data space associated with trajectories for which the event $E$ occurs. The normalization factor is given by the total volume of the physically allowed region, 
\begin{equation} 
\label{4.10} 
\mathcal D = \int_{\phi_{\rm min}}^{\phi_{\rm max}} d\omega. 
\end{equation} 
The criterion adopted in this work for successful inflation is that the inflationary epoch produces at least $60$ e-folds, $N_{\text{inf}}\gtrsim60$. For the chaotic potential,  \cite{Ashtekar:2011rm} showed that this requirement restricts the bounce value of the scalar field to the intervals 
\begin{equation} 
\label{4.10bb} 
\phi_B \in \left(-\phi_{\rm min},\phi_1\right) \cup \left(\phi_2,\phi_{\rm max}\right), 
\end{equation} 
where $\phi_{1, 2}$ are determined by 
\bq
\lb{eq3.36}
N_{\text{inf}}(\phi_i) \approx 60\; (i = 1, 2), 
\eq
so that $N_{\text{inf}}(\phi_1 < \phi_B < \phi_2) < 60$. For each value of $m$, the values of $\phi_{1, 2}$ 
are given in 
Table \ref{TableI} for the chaotic potential $\alpha_1 = \alpha_2 = 0$.  
Then, the probability of unsuccessful inflation is given by 
\begin{equation} 
\label{4.10cc} 
P\!\left(\mathrm{not\ realized}\right) \lesssim \frac{ \displaystyle \int_{\phi_1}^{\phi_2} d\omega }{ \displaystyle \int_{\phi_{\rm min}}^{\phi_{\rm max}} d\omega } \lesssim {\cal{O}}\left(\times10^{-6}\right), \end{equation} 
as shown in Table \ref{TableI}. This is comparable to the result obtained in \cite{Ashtekar:2011rm} for the WMAP
observation.

This result illustrates the well-known conclusion that, within the quadratic model, trajectories leading to sufficient slow-roll inflation occupy nearly the entire physically allowed phase space. For the various observational inputs considered in this work, the corresponding probabilities are summarized in Table~I, where one again finds that successful inflation remains overwhelmingly favored.

%%%%%%%%%%%%%%%%%%%%%%%%Table 2  -  Polynomial No Ekpy %%%%%%%%%%%%%%%%%%%%%%%%%%%%%%%%%%%%%%%
%%%%%%%%%%%%%%%%%%%%%%%%%%%%%%%%%%%%%%%%%%%%%%%%%%%%%%%%%%%%%%%%%%%%%%%%%%%%
\begin{table*}[t]
\centering
\begin{tabular}{lccrccccc}
                          & $m$                  & $H_*$                & $\phi_*$ & $\phi_{\text{min}}$ & $\phi_{\text{max}}$ & $\phi_1$ &$\phi_2$ & $P$     \\ \hline \hline
\multirow{2}{*}{Planck}   & $1.56\times 10^{-6}$ & $5.54\times 10^{-6}$ & $2.52$                        & $-557.10$& $571.15$& $-5.68$& $0.21$& $5.76\times 10^{-3}$ \\ 
                          & $6.93\times 10^{-7}$ & $8.73\times 10^{-6}$ & $-3.79$                       & $-732.21$& $746.25$& $-5.82$& $0.10$& $4.55\times 10^{-3}$\\ \hline
\multirow{2}{*}{ACT}      & $1.49\times 10^{-6}$ & $5.37\times 10^{-6}$ & $2.58$                        & $-565.49$& $579.54$& $-5.72$                       & $0.19$                        & $5.67\times 10^{-3}$ \\ 
                          & $6.51\times 10^{-7}$ & $8.57\times 10^{-6}$ & $-3.90$                       & $-747.84$& $761.88$& $-5.86$                       & $0.08$                        & $5.33\times 10^{-3}$ \\ \hline
\multirow{2}{*}{Combined} & $1.11\times 10^{-6}$ & $4.30\times 10^{-6}$ & $2.92$                        & $-624.48$& $638.52$& $-5.76$                       & $0.16$                        & $5.14\times 10^{-3}$ \\ 
                          & $4.31\times 10^{-7}$ & $7.51\times 10^{-6}$ & $-4.70$                       & $-858.60$& $872.65$& $-5.90$                       & $0.02$                        & $3.75\times 10^{-3}$ \\ \toprule
\end{tabular}
\caption{The physical quantities ($H_*, \phi_*, m$) for a given set of ($A_s, n_s$) from the observations, Planck, ACT or Planck + ACT + LB2 (Combined),  and the corresponding values of $\phi_{\text{min}}$, $\phi_{\text{max}}$, $\phi_1$, $\phi_2$, and  $P_{\text{LQC}}$ for   $\dot{\phi}_B > 0$, and $\alpha_1 = 0.14$ and $\alpha_2 = 6.644\times 10^{-3}$ in mLQC-II.}
\label{TableII}
\end{table*}

%%%%%%%%%%  PE Table
%%%%%%%%%%%%%% TABLE 2

%%%%%%%%%%%%%%%%%%%%%%%%%%%%%%%%%%%%%%%%%%%%%%%%%%%%%%%%%%
%% Fig. 3

\begin{figure}
    \centering
    \includegraphics[width=0.85\linewidth]{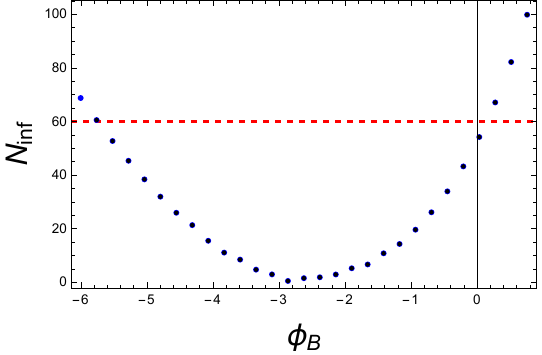}
    \caption{The plots of the total number of e-folds with a polynomial chaotic potential  given by Eq.\eqref{eq3.11} for $\alpha_1 = 0.14$, $\alpha_2 = 6.644\times 10^{-3}$ without an ekpyrotic potential in the framework of mLQC-II for the choice $\dot{\phi}_B>0$.  This case plots the inflation for $m=1.11\times10^{-6}m_P$ from the Combined data which can be found on Table \ref{TableII}.}
    \label{fig3}
\end{figure}
%%%%%%%%%%%%%%%%%%%%%%%%%%%%%%%%%%%%%%%%%%%%%%%%%%%%%%%%%%%%%

For the case of polynomial chaotic inflation of Eq.\eqref{eq3.11}, we consider the values $\alpha_1=0.14$ and $\alpha_2=6.644\times10^{-3}$.  Applying the same procedure as before we obtain the mass $m$, and can find the total e-fold defined in Eq.\eqref{efolds definition}.  Under the requirements that $N_{\text{inf}}\gtrsim60$, we find that the $\phi_B$ ranges must be given by Eq.(\ref{4.10bb}),
where $(\phi_1,\phi_2, \phi_{\text{min}},\phi_{\text{max}})$ can be found in Table \ref{TableII} for each given value of $m$.  Note that the multiple values of $m$ for each given set of $(A_s, n_s)$ are due to the fact that now Eqs.(\ref{eqA.38aa})-(\ref{eqA.40aa}) are higher-order  polynomial of the mass parameter $m$.  
In Fig. \ref{fig3} we plot $N_{\text{inf}}(\phi_B)$ for $\dot\phi_B > 0$. A similar plot for $\dot\phi_B < 0$ can also be obtained.
We next calculate plot $P\!\left(\mathrm{not\ realized}\right)$ and find  
\begin{equation} 
\label{4.10dd} 
P\!\left(\mathrm{not\ realized}\right) \lesssim \frac{ \displaystyle \int_{\phi_1}^{\phi_2} d\omega }{ \displaystyle \int_{\phi_{\rm min}}^{\phi_{\rm max}} d\omega } \lesssim {\cal{O}}\left(\times10^{-3}\right), \end{equation} 
as shown in Table \ref{TableII}. This is three orders of magnitude larger than the probabilities found in the chaotic potential case as shown in Table \ref{TableI}, but is quite comparable to those found in LQC and mLQC-I for the same polynomial chaotic potential \cite{Brown:2025hcb}.

\subsection{Effects of the Ekpyrotic Mechanism}

We begin with the parameter choice $U_0 = 0.0366$, $p = 0.1$, $\beta = 5$, originally considered in  \cite{McNamara:2022dmf} for the ekpyrotic potential. For the inflationary potential, we first consider the case $\alpha_1=\alpha_2=0$. Solving Eqs.\eqref{eqA.38aa}-\eqref{eqA.40aa} using the observational inputs $(A_s,n_s)$ yields the horizon-crossing quantities $(H_*,\phi_*,m)$. The resulting values for the Planck, ACT, and combined data sets are listed in Table \ref{TableIII}, from which we can see that there exists a uniqe value of $m$ for each given set of $(A_s, n_s)$, quite similar to the corresponding case without the ekpyrotic potential, as shown in Table \ref{TableI}.  For each value of $m$, the lower bound $\phi_{\rm Bmin}$ is obtained from Eq.\eqref{eq3.6} with $w_{\rm Bmin}=1.001$, while the upper bound $\phi_{\rm Bmax}$ follows from Eq.\eqref{eq3.6a}. For all three observational data sets, we find that $\phi_{\text{min}}$ and 
$\phi_{\text{max}}$ are the same, as shown in Table \ref{TableIII}, that is, the allowed range of the initial conditions of $\phi_B$ is 
\begin{equation} 
\label{eq3.9aa} 
\phi_B \in (-0.20,\,0.04),
\end{equation} 
which  is dramatically smaller than that in the absence of the ekpyrotic mechanism, as can be seen from   Tables \ref{TableI} and \ref{TableIII}.
 In the pure inflationary model, the allowed field values are determined by Eq.\eqref{phiMN}, yielding the broad domain $\phi\in(\phi_{\rm min},\phi_{\rm max})$. Once the ekpyrotic mechanism is introduced, however, the requirement $w_B>1$ near the bounce means our constraint Eq.\eqref{eq3.6a} restricts the allowed initial data to a much narrower interval $\phi\in(\phi_{\rm Bmin},\phi_{\rm Bmax})$. 
 
\begin{figure}
    \centering
    \includegraphics[width=6.5cm]{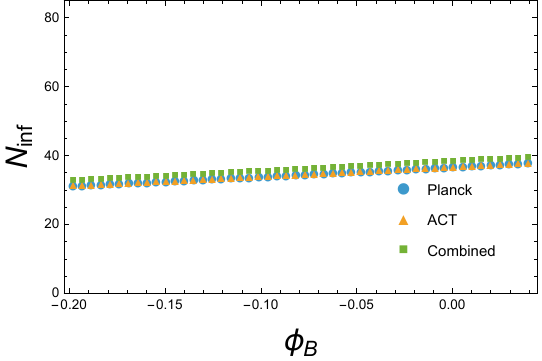}
    \caption{The plot of total number of e-folds with a chaotic potential, $\alpha_1 \alpha_2 = 0$, given by Eq.(\ref{eq3.11}), and an ekpyrotic potential with values $U_0=0.0366$. $p=0.1$, $\beta=5$, given by Eq.\eqref{eq3.3}.  The three cases on each graph correspond to different mass values given in Table \ref{TableIV}.}
    \label{fig4}
\end{figure}

 Using the masses listed in Table \ref{TableIII}, we numerically solve solutions across the entire interval $\phi_B\in(\phi_{\rm Bmin},\phi_{\rm Bmax})$ and compute the corresponding number of inflationary e-folds. The results are shown in Fig.\ref{fig4}. Unlike the case without an ekpyrotic phase, the duration of inflation is substantially reduced, with $N_{\rm inf} < 40$ throughout the allowed $\phi_B$ values given in Eq.(\ref{eq3.9aa}). This conclusion holds for other values of $m$ listed in  Table \ref{TableIII}. These results demonstrate that, for $U_0 = 0.0366$, $p = 0.1$, $\beta = 5$, the ekpyrotic sector substantially modifies the subsequent inflationary dynamics. Although it successfully generates an ekpyrotic phase around the bounce, it simultaneously reduces the duration of slow-roll inflation to $N_{\rm inf} < 40$, well below that required to resolve the standard cosmological puzzles.
 
%%%%%%%%%Table III    TableIII - Chaotic + Ekpyrotic
 %%%%%%%%%%%%%%%%%%%%%%%%%%%%%%%%%%%%%%%%%%%%%%%%%%
 %%%%%%%%%%%%%%%%%%%%%%%%%%%%%%%%%%%%%%%%%%%%%%%%%%
\begin{table*}[t]
\centering
\begin{tabular}{lcccccc}
         & $m$                  & $H_*$                & $\phi_*$ & $\phi_{\text{min}}$   & $\phi_{\text{max}}$  & $P$\\ \hline \hline
Planck   & $1.23\times 10^{-6}$ & $7.61\times 10^{-6}$ & $3.01$   & $-0.20$ & $0.04$&                  $1.00$\\ \hline 
ACT      & $1.18\times10^{-6}$  & $7.45\times 10^{-6}$ & $3.09$   & $-0.20$& $0.04$&                  $1.00$\\ \hline
Combined & $8.80\times 10^{-7}$ & $6.45\times10^{-6}$    & $3.58$   & $-0.20$& $0.04$&                 
$1.00$\end{tabular}
\caption{The physical quantities $(H_*,\phi_*,m)$ for a given set of $(A_s,n_s)$ from the observations, Planck, ACT or Planck + ACT + LB2 (Combined), and the corresponding $\phi_{\text{min}}$, $\phi_{\text{max}}$, and $P$ for $\dot{\phi}>0$ in a chaotic potential, $\alpha_1\alpha_2=0$ from Eq.\eqref{eq3.11}, with an ekpyrotic potential with values $U_0=0.0366$, $p=0.1$, $\beta=5$ from Eq.\eqref{eq3.4}.}
\label{TableIII}
\end{table*}

%%%%%%%%%Table III   TableIII
%%%%%%%%%%%%%%%%%%%%%%%%%%%%%%%%%%%%%%%%%%%%%%%%%%
%%%%%%%%%%%%%%%%%%%%%%%%%%%%%%%%%%%%%%%%%%%%%%%%%%

%%%%%%%%%%%%%%%%%  Figure for Chatic + Ekpyrotic Singh Parameters
%%%%%%%%%%%%%%%%%%%%%%%%%%%%%%%%%%%%%%%%%%%%%%%%%%

%%%%%%%%%%%%%%%%%  Figure for Chatoic + Ekpyrotic Matching Previous LQC parameters
\begin{figure}
    \centering
    \includegraphics[width=6.5cm]{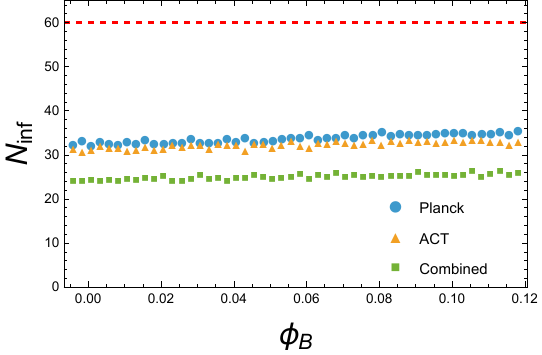}
    \caption{A plot of the number of e-folds for a chaotic potential , $\alpha_1 \alpha_2 = 0$, given by Eq.\eqref{eq3.11} plus an ekpyrotic potential for the values of $U_0=10^{-3}$, $p=10^{-2}$, $\beta=0.1$, given by Eq.\eqref{eq3.3}.  This plot demonstrates that at the same values of  $U_0$, $p$, and $\beta$ we no longer have proper inflation in mLQC-II when compared to LQC in \cite{Brown:2025hcb}.}
    \label{fig5}
\end{figure}
%%%%%%%%%%%%%%%%%%%%%%%%%%%%%%%%%%%%%%%%%%%%%%%%%%
%%%%%%%%%%%%%%%%%%%%%%%%%%%%%%%%%%%%%%%%%%%%%%%%%%

We next consider a different choice of ekpyrotic-potential parameters, $U_0 = 10^{-3}$, $p = 10^{-2}$, $\beta = 0.1$.  Ref. \cite{Brown:2025hcb} calculated the corresponding mass values and showed that, when combined with a chaotic inflationary potential, these values yield more than $60$ e-folds of inflation in standard LQC. We therefore examine the same parameter set in mLQC-II, again taking $\alpha_1=\alpha_2=0$. The resulting number of inflationary e-folds is displayed in Fig.\ref{fig5}. In contrast to the standard LQC, the inflationary phase in mLQC-II fails to generate the required duration of accelerated expansion. Throughout the allowed range of initial conditions, we find $N_{\rm inf} < 60$, indicating that the requirement for successful inflation is not satisfied.

\begin{figure}
    \centering
    \includegraphics[width=6.5cm]{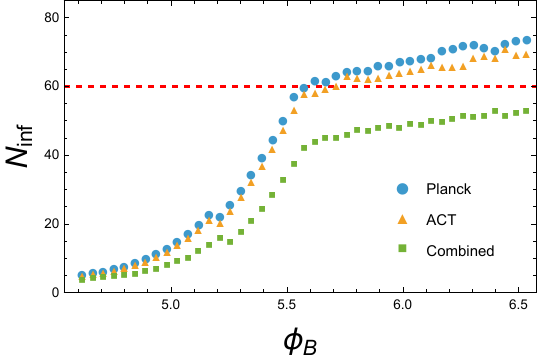}
    \caption{The plots of the total number of e-folds with a chaotic potential, $\alpha_1\alpha_2=0$ given by Eq.\eqref{eq3.11} and an ekpyrotic potential given by Eq.\eqref{eq3.3} in the framework of mLQC-II for the choice $\dot{\phi}_B>0$.  The choice of other parameters is $U_0=p=\beta =0.1$ and $m = 1.11 \times 10^{-6}\; m_P$ for  $w_B=1.0000001$. The three cases on each graph correspond to the three different masses given in Table \ref{TableIV}.}
    \label{fig6}
\end{figure}

However, for other choices of the free parameters in the theory,   we can realize e-folds larger than $60$.  For example,  consider the case for the ekpyrotic parameters $U_0=0.1$, $p=0.1$,$\beta=0.1$, while  lower the minimal value $W_{\text{Bmin}}$ to $w_B=1.0000001$.     We plot the number of e-folds in Fig.\ref{fig6}, and show that we surpass 60 e-folds.  The mass values, $\phi_{\text{min}}$, $\phi_{\text{max}}$, e-fold crossing $\phi_{60}$, and probability of not being realized for Planck, ACT, and the Combined are shown in  Table \ref{TableIV}. However, the probability for inflation to occur is very low
\begin{equation} 
\label{4.10ee} 
P\!\left(\mathrm{not\ realized}\right) \lesssim \frac{ \displaystyle \int_{\phi_1}^{\phi_2} d\omega }{ \displaystyle \int_{\phi_{\rm min}}^{\phi_{\rm max}} d\omega } \approx {\cal{O}}\left(1\right). \end{equation} 

%%%%%%%%%Table IV -  fig6
 %%%%%%%%%%%%%%%%%%%%%%%%%%%%%%%%%%%%%%%%%%%%%%%%%%
 %%%%%%%%%%%%%%%%%%%%%%%%%%%%%%%%%%%%%%%%%%%%%%%%%%
\begin{table*}[t]
\centering
\begin{tabular}{lccccccc}
         & $m$                  & $H_*$                & $\phi_*$ & $\phi_{\text{min}}$   & $\phi_{\text{max}}$   &$\phi_{60}$& $P$\\ \hline \hline
Planck   & $1.23\times 10^{-6}$ & $7.61\times 10^{-6}$ & $3.01$   & $-0.654$& $6.536$&$5.655$&                  $0.878$\\ \hline 
ACT      & $1.18\times10^{-6}$  & $7.45\times 10^{-6}$ & $3.09$   & $-0.654$& $6.536$&$5.802$&                  $0.898$\\ \hline
Combined & $8.80\times 10^{-7}$ & $6.45\times10^{-6}$    & $3.58$   & $-0.654$& $6.536$&N/A&                 
$1.000$\end{tabular}
\caption{The physical quantities $(H_*,\phi_*,m)$ for a given set of $(A_s,n_s)$ from the observations, Planck, ACT or Planck + ACT + LB2 (Combined), and the corresponding $\phi_{\text{min}}$, $\phi_{\text{max}}$, and $P$ for $\dot{\phi}>0$ in a chaotic inflationary potential, $\alpha_1\alpha_2=0$ given by Eq.\eqref{eq3.11} with an Ekpyrotic potential given by Eq.\eqref{eq3.3} with values $U_0=p=\beta=0.1$ for $w_B=1.0000001$.}
\label{TableIV}
\end{table*}

%%%%%%%%%%%%%%%%%%%%%%%%%%%%%%%%%%%%%%%%%%%%%%%%%%
%%%%%%%%%%%%%%%%%%%%%%%%%%%%%%%%%%%%%%%%%%%%%%%%%%
%%%%%%%%%%%%%%%% fig. 6
%%%%%%%%%%%%%%%%%%%%%%%%%%%%%%%%%%%%%%%%%%%%%%%%%%

More generally, our numerical analysis suggests that the inclusion of an ekpyrotic sector in mLQC-II significantly suppresses the duration of the subsequent inflationary phase. For the parameter choices investigated in this work, the number of inflationary e-folds typically falls within the range $N_{\rm inf}\sim 30-40$, well below the value usually required to resolve the standard cosmological observations.

\begin{figure}
    \centering
    \includegraphics[width=0.45\linewidth]{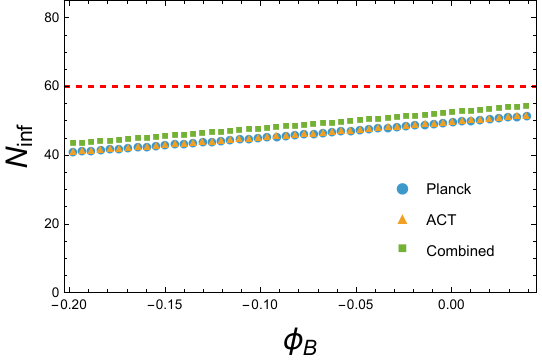}
    \caption{The plots of the total number of e-folds with a polynomial chaotic potential,  $\alpha_1=0.14$ and $\alpha_2=6.644\times10^{-3}$   given by Eq.\eqref{eq3.11} and an ekpyrotic potential given by Eq.\eqref{eq3.3} in the framework of mLQC-II for the choice $\dot{\phi}_B>0$.  The choice of parameters were chosen as $U_0=0.0366$, $p=0.1$, and $\beta=5$. The three cases on each graph correspond to the three different masses found in \cite{Brown:2025hcb}.}
    \label{PE-0366-01-5Plot}
\end{figure}

When we consider $\alpha_1\alpha_2\neq0$, we consider  we consider the first set of parameters $U_0=0.0366$, $p=0.1$, and $\beta=0.5$ along with $\alpha_1=0.14$ and $\alpha_2=6.644\times10^{-3}$.  Then, from Eqs.(\ref{eqA.38aa})-(\ref{eqA.40aa}) we find the possible mass parameters $m$ for each given set of $(A_s,n_s)$ displayed in Table \ref{???}
%TableV}.   
The total number of e-folds in the given range $(\phi_{min},\phi_{max})$ is always $N_{\text{inf}}<60$, as can be seen from Fig.\ref{PE-0366-01-5Plot}.  

\begin{figure}
    \centering
    \includegraphics[width=0.45\linewidth]{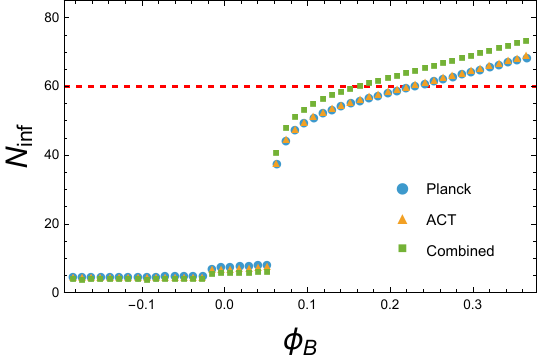}
    \caption{The plots of the total number of e-folds with a polynomial chaotic potential,  $\alpha_1=0.14$ and $\alpha_2=6.644\times10^{-3}$ given by Eq.\eqref{eq3.11} and an ekpyrotic potential given by Eq.\eqref{eq3.3} in the framework of mLQC-II for the choice $\dot{\phi}_B>0$.  The choice of parameters were chosen as $U_0=100$, $p=0.011$, and $\beta =0.5$. The three cases on each graph correspond to the three different masses given in Table \ref{TableV}.}
    \label{fig:placeholder}
\end{figure}

We then adjust the ekpyrotic potential in order to find more favorable parameters that yield a larger total number of e-folds, for example, choosing $U_0=100$, $p=0.011$ and $\beta=0.5$. In 
Table \ref{TableV} we show the physical quantities for such choices, while
in Fig.\ref{fig:placeholder} we plot $N_{\text{inf}}$ vs $\phi_B$, from which we can see that Planck and ACT give almost the exact same results for any given $\phi_B$ value.  For these cases, we find that both $\phi_{60}$ corresponds to $0.2312$ while the combined plot  yields $\phi_{60}\simeq0.1753$.  

%%%%%%%%%Table V - PE
%%%%%%%%%%%%%%%%%%%%%%%%%%%%%%%%%%%%%%%%%%%%%%%%%%
%%%%%%%%%%%%%%%%%%%%%%%%%%%%%%%%%%%%%%%%%%%%%%%%%%
\begin{table*}[t]
\centering
\begin{tabular}{lccccccc}
         & $m$                  & $H_*$                & $\phi_*$ & $\phi_{\text{min}}$   & $\phi_{\text{max}}$   &$\phi_{60}$& $P$\\ \hline \hline
Planck   & $6.93\times 10^{-7}$ & $8.73\times 10^{-6}$ & $-3.79$   & $-0.183$& $0.365$&$0.231$&                  $0.887$\\ \hline 
ACT      & $6.51\times10^{-7}$  & $8.57\times 10^{-6}$ & $-3.90$   & $-0.183$& $0.365$&$0.231$&                  $0.887$\\ \hline
Combined & $4.32\times 10^{-7}$ & $7.51\times10^{-6}$    & $-4.70$   & $-0.183$& $0.175$&N/A&                 
$0.838$\end{tabular}
\caption{The physical quantities $(H_*,\phi_*,m)$ for a given set of $(A_s,n_s)$ from the observations, Planck, ACT or Planck + ACT + LB2 (Combined), and the corresponding $\phi_{\text{min}}$, $\phi_{\text{max}}$, and $P$ for $\dot{\phi}>0$ in a polynomial chaotic potential,  $\alpha_1=0.14$ and $\alpha_2=6.644\times10^{-3}$ with an ekpyrotic potential with values $U_0=100$, $p=0.011$, and $\beta =0.5$.}
\label{TableV}
\end{table*}
%%%%%%%%%Table V - PE
%%%%%%%%%%%%%%%%%%%%%%%%%%%%%%%%%%%%%%%%%%%%%%%%%%
%%%%%%%%%%%%%%%%%%%%%%%%%%%%%%%%%%%%%%%%%%%%%%%%%%

We find that the probability of inflation not being realized is then $P(\text{not realized})\simeq88.7\%$ for Planck and ACT, and $P(\text{not realized})\simeq83.8\%$ for the combined case.  When compared to inflation without an ekpyrotic mechanism, we can see that the probabilities for inflation have significantly decreased and can no longer be considered generic. Therefore, for initial conditions to exist with $N_{\text{inf}}\gtrsim60$, fine-tuning of the parameters will be required.

\section{Conclusions and Remarks}

In this work, we investigated the effects of an ekpyrotic mechanism on the post-bounce inflationary dynamics of mLQC-II. We considered a single scalar field whose potential contains both an ekpyrotic contribution, which becomes negative near the quantum bounce, and an inflationary contribution that can dominate during the subsequent expanding phase. By requiring $w_B>1$, the scalar field grows more rapidly than the anisotropic shear during contraction, thereby providing a mechanism for suppressing the shear in the bounce region. Our numerical analysis shows, however, that the same ekpyrotic contribution can substantially alter the post-bounce evolution and reduce the duration of slow-roll inflation. For both the chaotic and polynomial inflationary potentials, we determined the inflationary mass parameter from the observational values of $(A_s,n_s)$ reported by Planck 2018, ACT 2025, and the combined data set. In the absence of the ekpyrotic contribution, sufficient inflation is overwhelmingly favored for the quadratic potential and remains highly probable for the polynomial potential. 

Once the ekpyrotic potential is included, the range of admissible initial conditions becomes significantly smaller. For several representative parameter choices, including $U_0=0.0366$, $p=0.1$, and $\beta=5$, the resulting inflationary phase fails to produce the required $N_{\rm inf}\gtrsim60$ e-folds. Similar suppression occurs for parameter values that previously produced sufficient inflation in standard LQC, demonstrating that the corresponding conclusions do not carry over directly to mLQC-II. Nevertheless, our results also show that sufficient inflation is not completely excluded. By appropriately adjusting the ekpyrotic parameters, we find nonempty regions of the initial-data space that produce more than $60$ e-folds. These successful regions are generally much more restricted than in models without an ekpyrotic phase, and their probabilities depend sensitively on the potential parameters and observationally inferred mass. We therefore conclude that the ekpyrotic mechanism can successfully suppress the shear in mLQC-II while remaining compatible with a sufficiently long inflationary phase, but the simultaneous realization of both features might need fine-tuning of initial conditions. Similar to \cite{Brown:2025hcb}, since our current studies are only restricted to numerical ones, our conclusions is not definitive, and a more systematic exploration of the ekpyrotic parameter space, together with an analysis of cosmological perturbations, will be necessary to determine the full viability and observational consequences of this model.

\begin{acknowledgments}

 We would like to express our gratitude to Dr. G. Cleaver and Mr. C. Brown and Mr. B. Phillips  for valuable discussions.  A.W. is partially supported by the US NSF grant: PHY2308845.

\end{acknowledgments}

\bibliographystyle{apsrev4-1}
\bibliography{Inflation_Ekpyrotic_Potentials}

@article{Brown:2025hcb,
    author = "Brown, Christian and Fier, Jared and Phillips, Brian and Cleaver, Gerald and Wang, Anzhong",
    title = "{Effects of the ekpyrotic mechanism on inflationary phase in loop quantum cosmologies}",
    eprint = "2510.03907",
    archivePrefix = "arXiv",
    primaryClass = "gr-qc",
    doi = "10.1140/epjc/s10052-026-16275-x",
    journal = "Eur. Phys. J. C",
    volume = "86",
    number = "8",
    pages = "1006",
    year = "2026"
}

@article{DESI:2025zpo,
    author = "Abdul Karim, M. and others",
    collaboration = "DESI",
    title = "{DESI DR2 results. I. Baryon acoustic oscillations from the Lyman alpha forest}",
    eprint = "2503.14739",
    archivePrefix = "arXiv",
    primaryClass = "astro-ph.CO",
    reportNumber = "FERMILAB-PUB-25-0167-PPD",
    doi = "10.1103/2wwn-xjm5",
    journal = "Phys. Rev. D",
    volume = "112",
    number = "8",
    pages = "083514",
    year = "2025"
}

@article{DESI:2025zgx,
    author = "Abdul Karim, M. and others",
    collaboration = "DESI",
    title = "{DESI DR2 results. II. Measurements of baryon acoustic oscillations and cosmological constraints}",
    eprint = "2503.14738",
    archivePrefix = "arXiv",
    primaryClass = "astro-ph.CO",
    reportNumber = "FERMILAB-PUB-25-0169-PPD",
    doi = "10.1103/tr6y-kpc6",
    journal = "Phys. Rev. D",
    volume = "112",
    number = "8",
    pages = "083515",
    year = "2025"
}

@article{AtacamaCosmologyTelescope:2025blo,
    author = "Louis, Thibaut and others",
    collaboration = "Atacama Cosmology Telescope",
    title = "{The Atacama Cosmology Telescope: DR6 power spectra, likelihoods and {\ensuremath{\Lambda}}CDM parameters}",
    eprint = "2503.14452",
    archivePrefix = "arXiv",
    primaryClass = "astro-ph.CO",
    reportNumber = "FERMILAB-PUB-25-0071-PPD",
    doi = "10.1088/1475-7516/2025/11/062",
    journal = "JCAP",
    volume = "11",
    pages = "062",
    year = "2025"
}

@article{Yogesh:2024iip,
    author = "Yogesh and Li, Bao-Fei and Gangopadhyay, Mayukh R. and Wang, Anzhong",
    title = "{The Dynamics of Reheating in Loop Quantum Cosmology}",
    eprint = "2408.00316",
    archivePrefix = "arXiv",
    primaryClass = "gr-qc",
    month = "8",
    year = "2024"
}

@article{Kallosh:2025ijd,
    author = "Kallosh, Renata and Linde, Andrei",
    title = "{On the present status of inflationary cosmology}",
    eprint = "2505.13646",
    archivePrefix = "arXiv",
    primaryClass = "hep-th",
    doi = "10.1007/s10714-025-03470-6",
    journal = "Gen. Rel. Grav.",
    volume = "57",
    number = "10",
    pages = "135",
    year = "2025"
}

@article{Wilson-Ewing:2012lmx,
    author = "Wilson-Ewing, Edward",
    title = "{The Matter Bounce Scenario in Loop Quantum Cosmology}",
    eprint = "1211.6269",
    archivePrefix = "arXiv",
    primaryClass = "gr-qc",
    doi = "10.1088/1475-7516/2013/03/026",
    journal = "JCAP",
    volume = "03",
    pages = "026",
    year = "2013"
}

@article{Wilson-Ewing:2013bla,
    author = "Wilson-Ewing, Edward",
    title = "{Ekpyrotic loop quantum cosmology}",
    eprint = "1306.6582",
    archivePrefix = "arXiv",
    primaryClass = "gr-qc",
    doi = "10.1088/1475-7516/2013/08/015",
    journal = "JCAP",
    volume = "08",
    pages = "015",
    year = "2013"
}

@article{Wilson-Ewing:2015sfx,
    author = "Wilson-Ewing, Edward",
    title = "{Separate universes in loop quantum cosmology: framework and applications}",
    eprint = "1512.05743",
    archivePrefix = "arXiv",
    primaryClass = "gr-qc",
    doi = "10.1142/S0218271816420025",
    journal = "Int. J. Mod. Phys. D",
    volume = "25",
    number = "08",
    pages = "1642002",
    year = "2016"
}

@article{Bojowald:2004kt,
    author = "Bojowald, Martin and Maartens, Roy and Singh, Parampreet",
    title = "{Loop quantum gravity and the cyclic universe}",
    eprint = "hep-th/0407115",
    archivePrefix = "arXiv",
    reportNumber = "AEI-2004-051",
    doi = "10.1103/PhysRevD.70.083517",
    journal = "Phys. Rev. D",
    volume = "70",
    pages = "083517",
    year = "2004"
}

@article{Li:2021fmu,
    author = "Li, Bao-Fei and Singh, Parampreet",
    title = "{Loop quantum gravity effects might restrict a cyclic evolution}",
    eprint = "2108.12553",
    archivePrefix = "arXiv",
    primaryClass = "gr-qc",
    doi = "10.1103/PhysRevD.105.046013",
    journal = "Phys. Rev. D",
    volume = "105",
    number = "4",
    pages = "046013",
    year = "2022"
}

@article{Li:2020pww,
    author = "Li, Bao-Fei and Saini, Sahil and Singh, Parampreet",
    title = "{Primordial power spectrum from a matter-Ekpyrotic bounce scenario in loop quantum cosmology}",
    eprint = "2012.10462",
    archivePrefix = "arXiv",
    primaryClass = "gr-qc",
    doi = "10.1103/PhysRevD.103.066020",
    journal = "Phys. Rev. D",
    volume = "103",
    number = "6",
    pages = "066020",
    year = "2021"
}

@article{Wands:1998yp,
    author = "Wands, David",
    title = "{Duality invariance of cosmological perturbation spectra}",
    eprint = "gr-qc/9809062",
    archivePrefix = "arXiv",
    reportNumber = "PU-RCG-98-15",
    doi = "10.1103/PhysRevD.60.023507",
    journal = "Phys. Rev. D",
    volume = "60",
    pages = "023507",
    year = "1999"
}

@article{Itzhaki:2025gdv,
    author = "Itzhaki, Nissan and Peleg, Uri and Steinhardt, Paul J.",
    title = "{Instant Folded Strings, Dark Energy and a Cyclic Bouncing Universe}",
    eprint = "2508.09745",
    archivePrefix = "arXiv",
    primaryClass = "gr-qc",
    month = "8",
    year = "2025"
}

@article{Ijjas:2024oqn,
    author = "Ijjas, Anna and Steinhardt, Paul J. and Garfinkle, David and Cook, William G.",
    title = "{Smoothing and flattening the universe through slow contraction versus inflation}",
    eprint = "2404.00867",
    archivePrefix = "arXiv",
    primaryClass = "gr-qc",
    doi = "10.1088/1475-7516/2024/07/077",
    journal = "JCAP",
    volume = "07",
    pages = "077",
    year = "2024"
}

@article{Tukhashvili:2023itb,
    author = "Tukhashvili, Giorgi and Steinhardt, Paul J.",
    title = "{Cosmological Bounces Induced by a Fermion Condensate}",
    eprint = "2307.16098",
    archivePrefix = "arXiv",
    primaryClass = "gr-qc",
    doi = "10.1103/PhysRevLett.131.091001",
    journal = "Phys. Rev. Lett.",
    volume = "131",
    number = "9",
    pages = "091001",
    year = "2023"
}

@article{Ijjas:2021zyf,
    author = "Ijjas, Anna and Pretorius, Frans and Steinhardt, Paul J. and Garfinkle, David",
    title = "{Dynamical attractors in contracting spacetimes dominated by kinetically coupled scalar fields}",
    eprint = "2109.09768",
    archivePrefix = "arXiv",
    primaryClass = "gr-qc",
    doi = "10.1088/1475-7516/2021/12/030",
    journal = "JCAP",
    volume = "12",
    number = "12",
    pages = "030",
    year = "2021"
}

@article{Ijjas:2020dws,
    author = "Ijjas, Anna and Cook, William G. and Pretorius, Frans and Steinhardt, Paul J. and Davies, Elliot Y.",
    title = "{Robustness of slow contraction to cosmic initial conditions}",
    eprint = "2006.04999",
    archivePrefix = "arXiv",
    primaryClass = "gr-qc",
    doi = "10.1088/1475-7516/2020/08/030",
    journal = "JCAP",
    volume = "08",
    pages = "030",
    year = "2020"
}

@article{Ijjas:2019pyf,
    author = "Ijjas, Anna and Steinhardt, Paul J.",
    title = "{A new kind of cyclic universe}",
    eprint = "1904.08022",
    archivePrefix = "arXiv",
    primaryClass = "gr-qc",
    doi = "10.1016/j.physletb.2019.06.056",
    journal = "Phys. Lett. B",
    volume = "795",
    pages = "666--672",
    year = "2019"
}

@article{Ashtekar:2011rm,
    author = "Ashtekar, Abhay and Sloan, David",
    title = "{Probability of Inflation in Loop Quantum Cosmology}",
    eprint = "1103.2475",
    archivePrefix = "arXiv",
    primaryClass = "gr-qc",
    reportNumber = "IGC-11-03-02",
    doi = "10.1007/s10714-011-1246-y",
    journal = "Gen. Rel. Grav.",
    volume = "43",
    pages = "3619--3655",
    year = "2011"
}

@article{Battefeld:2014uga,
    author = "Battefeld, D. and Peter, Patrick",
    title = "{A Critical Review of Classical Bouncing Cosmologies}",
    eprint = "1406.2790",
    archivePrefix = "arXiv",
    primaryClass = "astro-ph.CO",
    doi = "10.1016/j.physrep.2014.12.004",
    journal = "Phys. Rept.",
    volume = "571",
    pages = "1--66",
    year = "2015"
}

@article{Cai:2012va,
    author = "Cai, Yi-Fu and Easson, Damien A. and Brandenberger, Robert",
    title = "{Towards a Nonsingular Bouncing Cosmology}",
    eprint = "1206.2382",
    archivePrefix = "arXiv",
    primaryClass = "hep-th",
    doi = "10.1088/1475-7516/2012/08/020",
    journal = "JCAP",
    volume = "08",
    pages = "020",
    year = "2012"
}

@article{Ashtekar:2009vc,
    author = "Ashtekar, Abhay and Wilson-Ewing, Edward",
    title = "{Loop quantum cosmology of Bianchi I models}",
    eprint = "0903.3397",
    archivePrefix = "arXiv",
    primaryClass = "gr-qc",
    doi = "10.1103/PhysRevD.79.083535",
    journal = "Phys. Rev. D",
    volume = "79",
    pages = "083535",
    year = "2009"
}

@article{Chiou:2007sp,
    author = "Chiou, Dah-Wei and Vandersloot, Kevin",
    title = "{The Behavior of non-linear anisotropies in bouncing Bianchi I models of loop quantum cosmology}",
    eprint = "0707.2548",
    archivePrefix = "arXiv",
    primaryClass = "gr-qc",
    reportNumber = "IGPG-07-5-1",
    doi = "10.1103/PhysRevD.76.084015",
    journal = "Phys. Rev. D",
    volume = "76",
    pages = "084015",
    year = "2007"
}

@article{Bunch:1978yq,
    author = "Bunch, T. S. and Davies, P. C. W.",
    title = "{Quantum Field Theory in de Sitter Space: Renormalization by Point Splitting}",
    doi = "10.1098/rspa.1978.0060",
    journal = "Proc. Roy. Soc. Lond. A",
    volume = "360",
    pages = "117--134",
    year = "1978"
}

@article{Borde:1993xh,
    author = "Borde, Arvind and Vilenkin, Alexander",
    title = "{Eternal inflation and the initial singularity}",
    eprint = "gr-qc/9312022",
    archivePrefix = "arXiv",
    doi = "10.1103/PhysRevLett.72.3305",
    journal = "Phys. Rev. Lett.",
    volume = "72",
    pages = "3305--3309",
    year = "1994"
}

@article{Borde:2001nh,
    author = "Borde, Arvind and Guth, Alan H. and Vilenkin, Alexander",
    title = "{Inflationary space-times are incompletein past directions}",
    eprint = "gr-qc/0110012",
    archivePrefix = "arXiv",
    reportNumber = "MIT-CTP-3183",
    doi = "10.1103/PhysRevLett.90.151301",
    journal = "Phys. Rev. Lett.",
    volume = "90",
    pages = "151301",
    year = "2003"
}

@article{Brandenberger:2012aj,
    author = "Brandenberger, Robert H. and Martin, Jerome",
    title = "{Trans-Planckian Issues for Inflationary Cosmology}",
    eprint = "1211.6753",
    archivePrefix = "arXiv",
    primaryClass = "astro-ph.CO",
    doi = "10.1088/0264-9381/30/11/113001",
    journal = "Class. Quant. Grav.",
    volume = "30",
    pages = "113001",
    year = "2013"
}

@inproceedings{Silverstein:2016ggb,
    author = "Silverstein, Eva",
    title = "{TASI lectures on cosmological observables and string theory}",
    booktitle = "{Theoretical Advanced Study Institute in Elementary Particle Physics}: {New Frontiers in Fields and Strings}",
    eprint = "1606.03640",
    archivePrefix = "arXiv",
    primaryClass = "hep-th",
    doi = "10.1142/9789813149441_0009",
    pages = "545--606",
    year = "2017"
}

@book{Baumann:2014nda,
    author = "Baumann, Daniel and McAllister, Liam",
    title = "{Inflation and String Theory}",
    eprint = "1404.2601",
    archivePrefix = "arXiv",
    primaryClass = "hep-th",
    doi = "10.1017/CBO9781316105733",
    isbn = "978-1-107-08969-3, 978-1-316-23718-2",
    publisher = "Cambridge University Press",
    series = "Cambridge Monographs on Mathematical Physics",
    month = "5",
    year = "2015"
}

@article{Motaharfar:2023hil,
    author = "Motaharfar, Meysam and Singh, Parampreet and Thareja, Eklavya",
    title = "{Classicality and uniqueness in the loop quantization of Bianchi I spacetimes}",
    eprint = "2311.08465",
    archivePrefix = "arXiv",
    primaryClass = "gr-qc",
    doi = "10.1103/PhysRevD.109.086013",
    journal = "Phys. Rev. D",
    volume = "109",
    number = "8",
    pages = "086013",
    year = "2024"
}

@article{Khoury:2001wf,
    author = "Khoury, Justin and Ovrut, Burt A. and Steinhardt, Paul J. and Turok, Neil",
    title = "{The Ekpyrotic universe: Colliding branes and the origin of the hot big bang}",
    eprint = "hep-th/0103239",
    archivePrefix = "arXiv",
    doi = "10.1103/PhysRevD.64.123522",
    journal = "Phys. Rev. D",
    volume = "64",
    pages = "123522",
    year = "2001"
}

@article{Brandenberger:2016vhg,
    author = "Brandenberger, Robert and Peter, Patrick",
    title = "{Bouncing Cosmologies: Progress and Problems}",
    eprint = "1603.05834",
    archivePrefix = "arXiv",
    primaryClass = "hep-th",
    doi = "10.1007/s10701-016-0057-0",
    journal = "Found. Phys.",
    volume = "47",
    number = "6",
    pages = "797--850",
    year = "2017"
}

@article{Lehners:2008vx,
    author = "Lehners, Jean-Luc",
    title = "{Ekpyrotic and Cyclic Cosmology}",
    eprint = "0806.1245",
    archivePrefix = "arXiv",
    primaryClass = "astro-ph",
    doi = "10.1016/j.physrep.2008.06.001",
    journal = "Phys. Rept.",
    volume = "465",
    pages = "223--263",
    year = "2008"
}

@book{Ryan:1975jw,
    author = "Ryan, Michael P. and Shepley, Lawrence C.",
    title = "{Homogeneous Relativistic Cosmologies}",
    isbn = "978-0-691-08146-5, ",
    publisher = "Princeton University Press",
    address = "Princeton",
    series = "Princeton Series in Physics",
    year = "1975"
}

@article{Ashtekar:2011ni,
    author = "Ashtekar, Abhay and Singh, Parampreet",
    title = "{Loop Quantum Cosmology: A Status Report}",
    eprint = "1108.0893",
    archivePrefix = "arXiv",
    primaryClass = "gr-qc",
    doi = "10.1088/0264-9381/28/21/213001",
    journal = "Class. Quant. Grav.",
    volume = "28",
    pages = "213001",
    year = "2011"
}

@article{Li:2023dwy,
    author = "Li, Bao-Fei and Singh, Parampreet",
    title = "{Loop Quantum Cosmology: Physics of Singularity Resolution and its Implications}",
    eprint = "2304.05426",
    archivePrefix = "arXiv",
    primaryClass = "gr-qc",
    month = "4",
    year = "2023"
}

@article{Agullo:2023rqq,
    author = "Agull\'o, Ivan and Wang, Anzhong and Wilson-Ewing, Edward",
    title = "{Loop quantum cosmology: relation between theory and observations}",
    eprint = "2301.10215",
    archivePrefix = "arXiv",
    primaryClass = "gr-qc",
    month = "1",
    year = "2023"
}

@article{Li:2021mop,
    author = "Li, Bao-Fei and Singh, Parampreet and Wang, Anzhong",
    title = "{Phenomenological implications of modified loop cosmologies: an overview}",
    eprint = "2105.14067",
    archivePrefix = "arXiv",
    primaryClass = "gr-qc",
    doi = "10.3389/fspas.2021.701417",
    journal = "Front. Astron. Space Sci.",
    volume = "8",
    pages = "701417",
    year = "2021"
}

@article{Meissner:2004ju,
    author = "Meissner, Krzysztof A.",
    title = "{Black hole entropy in loop quantum gravity}",
    eprint = "gr-qc/0407052",
    archivePrefix = "arXiv",
    doi = "10.1088/0264-9381/21/22/015",
    journal = "Class. Quant. Grav.",
    volume = "21",
    pages = "5245--5252",
    year = "2004"
}

@article{Li:2018opr,
    author = "Li, Bao-Fei and Singh, Parampreet and Wang, Anzhong",
    title = "{Towards Cosmological Dynamics from Loop Quantum Gravity}",
    eprint = "1801.07313",
    archivePrefix = "arXiv",
    primaryClass = "gr-qc",
    doi = "10.1103/PhysRevD.97.084029",
    journal = "Phys. Rev. D",
    volume = "97",
    number = "8",
    pages = "084029",
    year = "2018"
}

@article{Li:2018fco,
    author = "Li, Bao-Fei and Singh, Parampreet and Wang, Anzhong",
    title = "{Qualitative dynamics and inflationary attractors in loop cosmology}",
    eprint = "1807.05236",
    archivePrefix = "arXiv",
    primaryClass = "gr-qc",
    doi = "10.1103/PhysRevD.98.066016",
    journal = "Phys. Rev. D",
    volume = "98",
    number = "6",
    pages = "066016",
    year = "2018"
}

@article{Li:2019ipm,
    author = "Li, Bao-Fei and Singh, Parampreet and Wang, Anzhong",
    title = "{Genericness of pre-inflationary dynamics and probability of the desired slow-roll inflation in modified loop quantum cosmologies}",
    eprint = "1906.01001",
    archivePrefix = "arXiv",
    primaryClass = "gr-qc",
    doi = "10.1103/PhysRevD.100.063513",
    journal = "Phys. Rev. D",
    volume = "100",
    number = "6",
    pages = "063513",
    year = "2019"
}

@article{Ashtekar:2004eh,
    author = "Ashtekar, Abhay and Lewandowski, Jerzy",
    title = "{Background independent quantum gravity: A Status report}",
    eprint = "gr-qc/0404018",
    archivePrefix = "arXiv",
    doi = "10.1088/0264-9381/21/15/R01",
    journal = "Class. Quant. Grav.",
    volume = "21",
    pages = "R53",
    year = "2004"
}

@article{Yang:2009fp,
    author = "Yang, Jinsong and Ding, You and Ma, Yongge",
    title = "{Alternative quantization of the Hamiltonian in loop quantum cosmology II: Including the Lorentz term}",
    eprint = "0904.4379",
    archivePrefix = "arXiv",
    primaryClass = "gr-qc",
    reportNumber = "AEI-2009-042",
    doi = "10.1016/j.physletb.2009.10.072",
    journal = "Phys. Lett. B",
    volume = "682",
    pages = "1--7",
    year = "2009"
}

@book{Thiemann_2007, place={Cambridge}, series={Cambridge Monographs on Mathematical Physics}, title={Modern Canonical Quantum General Relativity}, publisher={Cambridge University Press}, author={Thiemann, Thomas}, year={2007}, collection={Cambridge Monographs on Mathematical Physics}}

@book{Bojowald_2010, place={Cambridge}, title={Canonical Gravity and Applications: Cosmology, Black Holes, and Quantum Gravity}, publisher={Cambridge University Press}, author={Bojowald, Martin}, year={2010}}

@book{Gambini:2011zz,
    author = "Gambini, Rodolfo and Pullin, Jorge",
    title = "{A first course in loop quantum gravity}",
    year = "2011",
    publisher = {Oxford University Press}
}

@book{Rovelli_Vidotto_2014, place={Cambridge}, title={Covariant Loop Quantum Gravity: An Elementary Introduction to Quantum Gravity and Spinfoam Theory}, publisher={Cambridge University Press}, author={Rovelli, Carlo and Vidotto, Francesca}, year={2014}}

@book{Ashtekar:2017yom,
    editor = "Ashtekar, Abhay and Pullin, Jorge",
    title = "{Loop Quantum Gravity}: {The First 30 Years}",
    doi = "10.1142/10445",
    isbn = "978-981-320-992-3, 978-981-322-001-0, 978-981-320-993-0",
    publisher = "World Scientific",
    series = "100 Years of General Relativity",
    volume = "4",
    year = "2017"
}

@article{Zhu:2017jew,
    author = "Zhu, Tao and Wang, Anzhong and Cleaver, Gerald and Kirsten, Klaus and Sheng, Qin",
    title = "{Pre-inflationary universe in loop quantum cosmology}",
    eprint = "1705.07544",
    archivePrefix = "arXiv",
    primaryClass = "gr-qc",
    doi = "10.1103/PhysRevD.96.083520",
    journal = "Phys. Rev. D",
    volume = "96",
    number = "8",
    pages = "083520",
    year = "2017"
}

@article{Planck:2018jri,
    author = "Akrami, Y. and others",
    collaboration = "Planck",
    title = "{Planck 2018 results. X. Constraints on inflation}",
    eprint = "1807.06211",
    archivePrefix = "arXiv",
    primaryClass = "astro-ph.CO",
    doi = "10.1051/0004-6361/201833887",
    journal = "Astron. Astrophys.",
    volume = "641",
    pages = "A10",
    year = "2020"
}

@inproceedings{Baumann:2009ds,
    author = "Baumann, Daniel",
    title = "{Inflation}",
    booktitle = "{Theoretical Advanced Study Institute in Elementary Particle Physics}: {Physics of the Large and the Small}",
    eprint = "0907.5424",
    archivePrefix = "arXiv",
    primaryClass = "hep-th",
    reportNumber = "TASI-2009",
    doi = "10.1142/9789814327183_0010",
    pages = "523--686",
    year = "2011"
}

@ARTICLE{1981PhRvD..23..347G,
       author = {{Guth}, Alan H.},
        title = "{Inflationary universe: A possible solution to the horizon and flatness problems}",
      journal = {\prd},
         year = 1981,
        month = jan,
       volume = {23},
       number = {2},
        pages = {347-356},
          doi = {10.1103/PhysRevD.23.347},
       adsurl = {https://ui.adsabs.harvard.edu/abs/1981PhRvD..23..347G}
}

@book{Green_Schwarz_Witten_2012, place={Cambridge}, series={Cambridge Monographs on Mathematical Physics}, title={Superstring Theory: 25th Anniversary Edition}, publisher={Cambridge University Press}, author={Green, Michael B. and Schwarz, John H. and Witten, Edward}, year={2012}, collection={Cambridge Monographs on Mathematical Physics}}

@book{Becker:2006dvp,
    author = "Becker, K. and Becker, M. and Schwarz, J. H.",
    title = "{String theory and M-theory: A modern introduction}",
    doi = "10.1017/CBO9780511816086",
    isbn = "978-0-511-25486-4, 978-0-521-86069-7, 978-0-511-81608-6",
    publisher = "Cambridge University Press",
    month = "12",
    year = "2006"
}

@book{Rovelli:2014ssa,
    author = "Rovelli, Carlo and Vidotto, Francesca",
    title = "{Covariant Loop Quantum Gravity}: {An Elementary Introduction to Quantum Gravity and Spinfoam Theory}",
    isbn = "978-1-107-06962-6, 978-1-316-14729-0",
    publisher = "Cambridge University Press",
    series = "Cambridge Monographs on Mathematical Physics",
    month = "11",
    year = "2014"
}

@article{ElizagaNavascues:2020uyf,
    author = "Elizaga Navascu\'es, Beatriz and Marug\'an, Guillermo A. Mena",
    title = "{Hybrid Loop Quantum Cosmology: An Overview}",
    eprint = "2011.04559",
    archivePrefix = "arXiv",
    primaryClass = "gr-qc",
    doi = "10.3389/fspas.2021.624824",
    journal = "Front. Astron. Space Sci.",
    volume = "8",
    pages = "81",
    year = "2021"
}

@article{Assanioussi:2018hee,
    author = "Assanioussi, Mehdi and Dapor, Andrea and Liegener, Klaus and Paw\l{}owski, Tomasz",
    title = "{Emergent de Sitter Epoch of the Quantum Cosmos from Loop Quantum Cosmology}",
    eprint = "1801.00768",
    archivePrefix = "arXiv",
    primaryClass = "gr-qc",
    doi = "10.1103/PhysRevLett.121.081303",
    journal = "Phys. Rev. Lett.",
    volume = "121",
    number = "8",
    pages = "081303",
    year = "2018"
}

@article{Assanioussi:2019iye,
    author = "Assanioussi, Mehdi and Dapor, Andrea and Liegener, Klaus and Paw\l{}owski, Tomasz",
    title = "{Emergent de Sitter epoch of the Loop Quantum Cosmos: a detailed analysis}",
    eprint = "1906.05315",
    archivePrefix = "arXiv",
    primaryClass = "gr-qc",
    doi = "10.1103/PhysRevD.100.084003",
    journal = "Phys. Rev. D",
    volume = "100",
    number = "8",
    pages = "084003",
    year = "2019"
}

@article{Dapor:2017rwv,
    author = "Dapor, Andrea and Liegener, Klaus",
    title = "{Cosmological Effective Hamiltonian from full Loop Quantum Gravity Dynamics}",
    eprint = "1706.09833",
    archivePrefix = "arXiv",
    primaryClass = "gr-qc",
    doi = "10.1016/j.physletb.2018.09.005",
    journal = "Phys. Lett. B",
    volume = "785",
    pages = "506--510",
    year = "2018"
}

@article{Dapor:2017gdk,
    author = "Dapor, Andrea and Liegener, Klaus",
    title = "{Cosmological coherent state expectation values in loop quantum gravity I. Isotropic kinematics}",
    eprint = "1710.04015",
    archivePrefix = "arXiv",
    primaryClass = "gr-qc",
    doi = "10.1088/1361-6382/aac4ba",
    journal = "Class. Quant. Grav.",
    volume = "35",
    number = "13",
    pages = "135011",
    year = "2018"
}

@article{Han:2021cwb,
    author = "Han, Muxin and Liu, Hongguang",
    title = "{Loop quantum gravity on dynamical lattice and improved cosmological effective dynamics with inflaton}",
    eprint = "2101.07659",
    archivePrefix = "arXiv",
    primaryClass = "gr-qc",
    doi = "10.1103/PhysRevD.104.024011",
    journal = "Phys. Rev. D",
    volume = "104",
    number = "2",
    pages = "024011",
    year = "2021"
}

@article{McNamara:2022dmf,
    author = "McNamara, A. Meenakshi and Saini, Sahil and Singh, Parampreet",
    title = "{Novel relationship between shear and energy density at the bounce in nonsingular Bianchi I spacetimes}",
    eprint = "2210.07257",
    archivePrefix = "arXiv",
    primaryClass = "gr-qc",
    doi = "10.1103/PhysRevD.107.026003",
    journal = "Phys. Rev. D",
    volume = "107",
    number = "2",
    pages = "026003",
    year = "2023"
}

@article{Destri:2007pv,
    author = "Destri, C. and de Vega, Hector J. and Sanchez, N. G.",
    title = "{MCMC analysis of WMAP3 and SDSS data points to broken symmetry inflaton potentials and provides a lower bound on the tensor to scalar ratio}",
    eprint = "astro-ph/0703417",
    archivePrefix = "arXiv",
    doi = "10.1103/PhysRevD.77.043509",
    journal = "Phys. Rev. D",
    volume = "77",
    pages = "043509",
    year = "2008"
}

@article{Nakayama:2013jka,
    author = "Nakayama, Kazunori and Takahashi, Fuminobu and Yanagida, Tsutomu T.",
    title = "{Polynomial Chaotic Inflation in the Planck Era}",
    eprint = "1303.7315",
    archivePrefix = "arXiv",
    primaryClass = "hep-ph",
    reportNumber = "TU-932, IPMU13-0070, UT-13-18",
    doi = "10.1016/j.physletb.2013.06.050",
    journal = "Phys. Lett. B",
    volume = "725",
    pages = "111--114",
    year = "2013"
}

@article{Kallosh:2014xwa,
    author = "Kallosh, Renata and Linde, Andrei and Westphal, Alexander",
    title = "{Chaotic Inflation in Supergravity after Planck and BICEP2}",
    eprint = "1405.0270",
    archivePrefix = "arXiv",
    primaryClass = "hep-th",
    reportNumber = "DESY-14-066",
    doi = "10.1103/PhysRevD.90.023534",
    journal = "Phys. Rev. D",
    volume = "90",
    number = "2",
    pages = "023534",
    year = "2014"
}

@article{ACT:2025fju,
    author = "Louis, Thibaut and others",
    collaboration = "ACT",
    title = "{The Atacama Cosmology Telescope: DR6 Power Spectra, Likelihoods and $\Lambda$CDM Parameters}",
    eprint = "2503.14452",
    archivePrefix = "arXiv",
    primaryClass = "astro-ph.CO",
    reportNumber = "FERMILAB-PUB-25-0071-PPD",
    month = "3",
    year = "2025"
}

\end{document}